\documentclass[lettersize,journal]{IEEEtran}
\IEEEoverridecommandlockouts
\usepackage{cite}
\usepackage{multirow}
\usepackage{amsmath,amssymb,amsfonts,amssymb}
\usepackage{amsthm}
\usepackage[table]{xcolor} 

\usepackage[table]{xcolor}

\newtheorem{theorem}{Theorem}
\newtheorem{lemma}[theorem]{Lemma}

\usepackage{algorithmic}
\usepackage{graphicx}
\usepackage{subcaption}
\usepackage{booktabs}
\usepackage{textcomp}
\usepackage{xcolor}
\usepackage{placeins}
\usepackage{bbm}
\usepackage{url}
\usepackage{soul} 
\def\BibTeX{{\rm B\kern-.05em{\sc i\kern-.025em b}\kern-.08em
    T\kern-.1667em\lower.7ex\hbox{E}\kern-.125emX}}
\usepackage[utf8]{inputenc}
\usepackage[T1]{fontenc}
\usepackage{cite}
\usepackage{amsmath,amssymb,amsfonts}
\usepackage{algorithmic}
\usepackage{graphicx}
\usepackage{textcomp}
\usepackage{xcolor}
\usepackage{url}
\usepackage[caption=false,font=footnotesize]{subfig}

\usepackage{booktabs}
\usepackage{multirow}
\usepackage{array}
\usepackage{makecell}
\usepackage{colortbl}

\usepackage{subcaption}
\usepackage{float}
\usepackage{placeins}

\usepackage{balance}
\usepackage{soul}

\begin{document}

\title{Filterless Multi-Color VLC via DC-Biased QCT}

\author{
\IEEEauthorblockN{
Idris Cinemre\IEEEauthorrefmark{1},
Serkan Vela\IEEEauthorrefmark{2},
Gokce Hacioglu\IEEEauthorrefmark{2}\thanks{Corresponding author: Idris Cinemre (e-mail: idris.1.cinemre@kcl.ac.uk).}
}\\
\IEEEauthorblockA{\IEEEauthorrefmark{1}Department of Engineering, King's College London, London, UK}\\
\IEEEauthorblockA{\IEEEauthorrefmark{2}Dept. of Electrical and Electronics Engineering, Karadeniz Technical University, Trabzon, Turkey}
}

\maketitle

\begin{abstract}
Multi-color visible light communication (VLC) can increase throughput and enable joint lighting and communication operation, but practical color-based schemes such as color shift keying (CSK) typically rely on receiver optical filters whose nonideal passbands and spectral overlap introduce color crosstalk and significant SNR loss. This paper proposes a DC-biased quartered composite transform (QCT) transmission framework for quadrichromatic red, amber, green, blue (RAGB) luminaires that enables filterless multiple streams reception with a single photodiode. The method partitions the information symbols into four parallel real-valued streams and applies a set of mutually orthogonal QCT synthesis matrices designed from the invariances of the matched-filtered circulant channel; at the receiver, matched filtering and QCT-domain projection yield four decoupled scalar subchannels that admit single-tap equalization. A unified evaluation is carried out under common illumination constraints (CCT/CRI and illuminance uniformity) and throughput-matched configurations against RAGB-CSK and conventional DCO-OFDM baselines. In an indoor scenario, QCT attains up to $48.95~\mathrm{dB}$ average effective SNR, providing $15.1-22.7~\mathrm{dB}$ gain over CSK and $15.6-26.4~\mathrm{dB}$ gain over DCO-OFDM, while achieving essentially identical BER to DCO-OFDM in linear AWGN. Under matched mean optical power, QCT also yields near zero clipping distortion and a consistent $0.7-1~\mathrm{dB}$ PAPR reduction relative to DCO-OFDM, supporting power efficient and robust filterless multi-color VLC without sacrificing lighting quality.
\end{abstract}


\section{Introduction}

Visible light communication (VLC) systems typically employ intensity modulation and direct detection (IM/DD), where light-emitting diodes (LEDs) act simultaneously as illumination sources and optical transmitters~\cite{7072557,mapunda2020indoor,vappangi2018low}. Practical high-speed indoor VLC must therefore meet communication objectives while respecting strict lighting requirements, including flicker-free operation, target correlated color temperature (CCT), sufficient color rendering index (CRI), and adequate luminous flux/illuminance~\cite{6852085}.

A fundamental bottleneck in many commercial luminaires is the front-end modulation bandwidth. Phosphor-converted white LEDs are widely adopted due to low cost and simple driving circuits~\cite{9712353}, but their IM response is inherently low-pass because of the phosphor relaxation time, yielding unequalized $3$-dB bandwidths typically on the order of a few MHz (e.g., $\sim 1-3~\mathrm{MHz}$~\cite{4547916, komine2004fundamental,kucuk2021performance}). This limitation can be partially alleviated by using a blue-pass optical filter at the receiver to suppress the phosphor-emitted component such that the faster blue-chip response dominates, thereby extending the effective bandwidth to the tens of MHz range (e.g., $\sim 20~\mathrm{MHz}$ ~\cite{6249713,grubor2007wireless}). This bandwidth gain, however, is achieved at the expense of reduced received optical power at the photodetector (PD),  which generally degrades the available SNR and energy efficiency unless compensated by higher transmit power or increased receiver sensitivity. Beyond optical filtering, the usable bandwidth of baseline single carrier IM/DD VLC waveforms (e.g., OOK) is commonly extended using equalization \cite{4547916, 8307242, 8970270, aravindan20211, wang2023transmitter} to compensate the low-pass front-end response and mitigate inter-symbol interference (ISI). However, at very high symbol rates, adequate suppression of ISI and mitigation of nonlinear distortion may require advanced adaptive equalizers with increased complexity and implementation burden \cite{li2024pre, mitra2016adaptive, li2016wireless}, motivating multicarrier alternatives that simplify channel compensation.
Accordingly, a widely adopted single channel strategy for increasing throughput under a bandwidth limited front-end is to employ spectrally efficient multicarrier modulation with digital equalisation. In particular, orthogonal frequency-division multiplexing (OFDM) has been extensively investigated for IM/DD VLC because it provides high spectral efficiency and mitigates ISI through low-complexity per-subcarrier frequency-domain equalisation~\cite{4785281, 6415964,armstrong2006power}. Nevertheless, the high peak-to-average power ratio (PAPR) of OFDM~\cite{9850376,a2024optical} and the limited LED linear dynamic range typically necessitate DC biasing and clipping control~\cite{4524234,elgala2010led}, which can reduce power efficiency~\cite{9623482}.

To scale the aggregate data rate beyond what is achievable within a single optical channel, multi-color luminaires (e.g., red, green, blue (RGB)) can be exploited via spectral multiplexing (wavelength-division multiplexing, WDM), where independent data streams are conveyed on different spectral components and combined with bandwidth efficient modulation/equalization per color to increase the achievable sum rate while maintaining illumination functionality \cite{cossu20123,wang20154,chi20163, 7454689}. However, the practical gain of multi-color transmission is contingent on effective color channel separation at the receiver, since LED emission spectra, detector responsivities, and realizable optical filter passbands are not perfectly orthogonal, leading to spectral/color crosstalk and loss of channel orthogonality \cite{mathias2021modeling}. The color separation at receiver is typically achieved either via optical filter banks with multiple photodiodes\cite{8546751, 9329045}, or by treating color mixing as a small MIMO problem and applying post-equalization\cite{pergoloni2018space} to mitigate inter-channel interference and recover each color streams.

From a modulation perspective, IEEE~802.15.7 PHY~III \cite{8697198} standardises a color space scheme based on color shift keying (CSK) for RGB transmitters, in which data are conveyed by mapping symbols to CIE~1931 chromaticity points \cite{smith1931cie} through controlled intensity ratios among the LED primaries while maintaining a desired white average emission~\cite{rajagopal2012ieee,campos2018constellation,6780585}. By operating with a constant optical power envelope per symbol, CSK enables flicker-free dimmable transmission with stable perceived color~\cite{singh2018coded,monteiro2014design,mejia2019coding,tran2019effective}. Standardised CSK constellation design typically targets fixed average CCT and CRI; however, meeting these color quality constraints can reduce the achievable average luminous flux for a given electrical power budget~\cite{rajagopal2012ieee}. Moreover, practical CSK implementations typically rely on optical color filters for primary separation, and nonideal filtering can introduce substantial crosstalk that degrades SINR and constrains attainable data rates~\cite{10.1109/jsen.2015.2453200,10.3390/electronics10030262,10.1109/access.2020.2976537}. 
Consequently, the receiver front-end limitations that affect WDM-based multi-color links also directly impact CSK-based systems. Beyond the trichromatic RGB case standardised in IEEE~802.15.7, quadrichromatic luminaires, often based on red, amber, green, and blue (RAGB)\cite{liang2017constellation} or yellow (RGBY) \cite{7296576} LEDs, further expand the modulation dimensionality from three to four channels~\cite{ahn2012color,singh2015higher}. This additional degree of freedom enables four dimensional signalling that supports higher order constellations and allows simultaneous optimisation of photometric and communication performance \cite{7296576}. In particular, RAGB (or more generally QLED) luminaires can offer broader color gamuts, higher CRI, and potentially improved luminous efficacy relative to standard RGB devices~\cite{liang2017constellation,10.1117/1.oe.59.5.055102,10.1002/jsid.764,10.1002/aelm.202100598}, making them attractive for joint illumination and high-speed data transmission.

The severity of the front-end bottleneck is further amplified under realistic deployment conditions. Color crosstalk arises not only from intrinsic spectral overlap between LED emissions and filter passbands, but is exacerbated by the angular dependence of thin film interference filters~\cite{kucuk2024optical,ge2019optical}. As the angle of incidence increases, filter passband blue shift induces spectral misalignment, which can cause substantial SINR degradation in mobile and wide field of view scenarios~\cite{kucuk2021performance}. In dispersive indoor channels, high-speed operation additionally introduces ISI; the joint presence of ISI and color crosstalk can create a challenging interference landscape and may produce irreducible error floors if not properly mitigated~\cite{singh2018coded}. Prior work has therefore investigated improved optical filtering (e.g., optimised multilayer thin film filters and optical filter bank receivers) and complementary DSP beyond the basic IEEE~802.15.7 color calibration inversion (e.g., frequency-domain equalisation, rate-adaptive coding, and adaptive time-domain equalisation)~\cite{kucuk2021performance,ge2019optical,10.1109/tcomm.2019.2904503,singh2018coded,10.3390/electronics10030262,10.1049/iet-com.2018.5324}. Nevertheless, many state of the art multi-color solutions remain dependent on bulky and costly optical filtering and continue to exhibit sensitivity to angle of incidence variations, spectral mismatch, and device nonidealities, thereby motivating filterless receiver architectures in which color separation and crosstalk mitigation are shifted from the optical domain to robust, low complexity electrical domain processing.

This paper develops a filterless multiple streams transmission framework for RAGB VLC based on the \emph{quartered composite transform (QCT)}. QCT, was originally introduced as a structured transform for IM/DD waveform synthesis and PAPR reduction~\cite{Cinemre2024QCTP}, employs mutually orthogonal transform matrices to generate multiple simultaneously transmitted waveforms with improved ISI robustness compared with conventional DCO-OFDM. Here, it is shown that QCT can be leveraged to enable digital color-domain separation with a single filterless PD, in which color separation and crosstalk mitigation are performed entirely in the electrical domain rather than via optical filtering. The key observation is that, under the widely used point source abstraction for colocated multichip packages~\cite{7138560}, the four RAGB primaries share a common baseband channel impulse response. After matched filtering, the resulting circulant channel operator admits commuting symmetry invariances that can be exploited to construct four orthogonal, channel-invariant subspaces. By mapping four data streams onto these subspaces through QCT synthesis matrices, the receiver can recover four interference-free streams via low-complexity linear projections followed by single-tap equalization without any optical filtering. In addition, by appropriately constraining the average drive levels of each LED, QCT can maintain stable CCT and CRI and avoid the reduction in average luminous flux that is typical of conventional CSK.

We consider VLC based on RAGB multi-chip LEDs and benchmark three throughput-matched transmission strategies under a common room/luminaire configuration and common color quality targets: (i) an RAGB-CSK baseline consistent with IEEE~802.15.7 color space signalling (with a filter bank receiver), (ii) a conventional single stream DCO-OFDM baseline in which a common OFDM waveform is applied across the RAGB primaries to preserve a fixed color mixture operating point, and (iii) the proposed DC-biased QCT-based scheme that transmits four decorrelated streams and is received by a single filterless PD. The QCT approach removes filter-induced crosstalk and reduces receiver hardware complexity, while also improving robustness to the DC-biasing/clipping tradeoff in IM/DD transmission.

The main contributions of this paper are summarized as follows:
\begin{itemize}

  \item \textbf{\textit{Symmetry-driven QCT construction and decoupling.}} By exploiting the invariances of the matched-filtered circulant channel operator, we construct mutually orthogonal QCT subspaces and show that inter-stream coupling vanishes, yielding per-stream diagonal effective channels and \emph{single-tap equalization} (derived in the appendices).
  
  \item \textbf{\textit{Filterless multistream RAGB transmission via QCT.}} We propose a DC-biased QCT transmission method that maps four parallel real-valued streams onto the RAGB primaries and enables recovery from a \emph{single filterless PD} using low-complexity linear processing.

  \item \textbf{\textit{Throughput-matched performance evaluation and nonlinearity analysis.}} We provide a comparative evaluation against RAGB-CSK and DCO-OFDM in an indoor room scenario, reporting BER and effective SNR/SINR, and we quantify IM/DD nonlinearity effects via clipping power/EVM and PAPR under matched operating points.
    \item \textbf{\textit{Unified RAGB VLC modeling with lighting metrics.}} We formulate a common system model covering RAGB-CSK, conventional DCO-OFDM, and the proposed QCT-based transmission, and integrate illumination and color quality metrics (illuminance/uniformity, CCT, and CRI) alongside communication metrics under a consistent simulation framework.

\end{itemize}

The remainder of this paper is organized as follows. Section~II presents the system models for CSK, DCO-OFDM, and the proposed QCT scheme. Section~III summarizes the illumination and color quality metrics used in the evaluation. Section~IV reports numerical results and comparisons. Section~V concludes the paper, and the appendices provide the single-tap equalization background and the QCT decoupling and power-moment derivations.

\textit{Notations:} Bold lower-case letters (e.g., $\mathbf{a}$) denote column vectors and bold upper-case letters (e.g., $\mathbf{A}$) denote matrices. The superscripts $(\cdot)^{\mathrm T}$ and $(\cdot)^{\mathrm H}$ represent transpose and conjugate transpose, respectively, while $(\cdot)^{*}$ denotes complex conjugation. The operator $\operatorname{flip}(\cdot)$ reverses the order of the elements of a vector. The symbol $\mathbf{I}_N$ denotes the $N\times N$ identity matrix, $\mathbf{0}_N$ denotes the length-$N$ zero vector, and $\mathbf{1}_n$ denotes the length-$n$ all-ones column vector. The indicator function $\mathbbm{1}_{\mathcal{A}}$ equals $1$ if the condition/event $\mathcal{A}$ holds and $0$ otherwise.

\section{SYSTEM MODEL}

\subsection{RAGB‑Based CSK Scheme}

\begin{figure}[t!]
\centering
\includegraphics[width=1.05\linewidth]{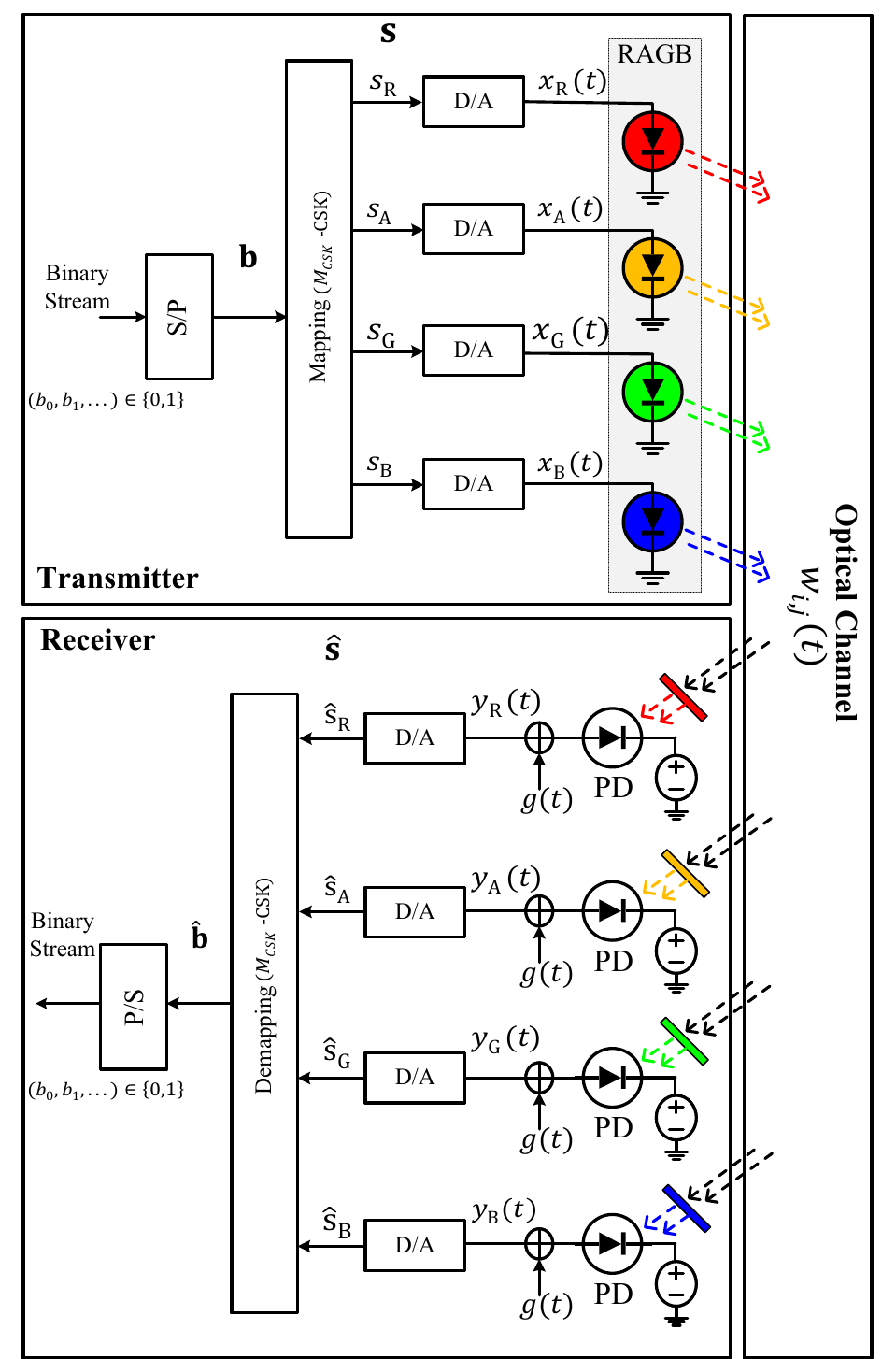}
\caption{RAGB‑based CSK transceiver with four channels}
\label{f:cskfig}
\end{figure}

In a four‑channel CSK system employing an RAGB LED array, as shown in Fig.~\ref{f:cskfig}, the input binary stream $(b_0,b_1,\ldots)\in\{0,1\}$ is segmented into words of $m_\mathrm{CSK}=\log_{2} M_\mathrm{CSK}$ bits. Each word
\begin{equation}
\mathbf{b}=\bigl[b_{0},\, b_{1},\,\ldots,\, b_{m_\mathrm{CSK}-1}\bigr]^{\mathrm T}\in\{0,1\}^{m_\mathrm{CSK}}
\end{equation}
is mapped by an $M_\mathrm{CSK}$‑ary CSK mapper to a four‑dimensional nonnegative optical intensity vector
\begin{equation}
\mathbf{s}=\bigl[s_{\mathrm{R}},\,s_{\mathrm{A}},\,s_{\mathrm{G}},\,s_{\mathrm{B}}\bigr]^{\mathrm T} \in\mathbb{R}_{+}^{4},
\end{equation}
whose components drive the red, amber, green, and blue LEDs, respectively. In standard CSK, the total emitted luminous flux per symbol is constrained to be constant,
\begin{equation}
\mathbf{1}_4^{\mathrm T}\mathbf{s} = \Phi,
\end{equation}
where $\mathbf{1}_4=[1,1,1,1]^{\mathrm T}$ and $\Phi$ is the prescribed optical power (proportional to the luminous flux), ensuring a constant perceived brightness.

After D/A conversion, the corresponding drive currents control the individual LEDs, and the superposed optical signal propagates through a free‑space VLC channel. At the receiver, the incoming light passes through an array of four optical color filters (red, amber, green, and blue) designed to physically separate the constituent color components before detection by corresponding PDs. Because the spectral power distributions (SPDs) of the LEDs overlap and the filters are non‑ideal, inter‑color interference (ICI) arises. The aggregate optical–electrical coupling is modelled by a nonnegative real matrix $\mathbf{W}\in\mathbb{R}_{+}^{4\times 4}$, whose diagonal elements $w_{i,i}$ denote the line‑of‑sight (LoS) gain from the $i$‑th LED emitter to the intended detection channel, while each off‑diagonal entry $w_{i,j}$ ($i\neq j$) measures crosstalk from the $j$‑th emitter into the $i$‑th channel. Here $i,j\in\{1,2,3,4\}$ correspond to the color channels $\{\mathrm{R},\mathrm{A},\mathrm{G},\mathrm{B}\}$.

The sampled received electrical vector $\mathbf{y} \in \mathbb{R}^4$ is therefore
\begin{equation}
\mathbf{y} \;=\; \mathbf{W}\mathbf{s} + \mathbf{g},
\end{equation}
where $\mathbf{g}\sim\mathcal N(\mathbf 0,\sigma_{g}^{2}\mathbf I_4)$ aggregates thermal and shot noise at the PD outputs.

Following A/D conversion, demodulation is performed via maximum-likelihood detection (MLD).
Let $\mathcal{S}_{\mathrm{CSK}}=\{\mathbf{s}_{\ell}\}_{\ell=0}^{M_{\mathrm{CSK}}-1}\subset\mathbb{R}_+^{4}$
denote the $M_{\mathrm{CSK}}$-ary CSK constellation. The detected index is
\begin{equation}
    \hat{\ell} = \arg\min_{\ell \in \{0, \dots, M_{\mathrm{CSK}}-1\}}
    \bigl\|\mathbf{y} - \mathbf{W}\mathbf{s}_{\ell} \bigr\|^{2}.
\end{equation}
The detected index $\hat{\ell}$ is finally demapped to the estimated bit word $\hat{\mathbf{b}}$, thereby reconstructing the transmitted
data stream.

\subsection{DC‑Biased Optical OFDM (DCO‑OFDM)}

\begin{figure}[t!]
  \centering
  \includegraphics[width=3.6 in]{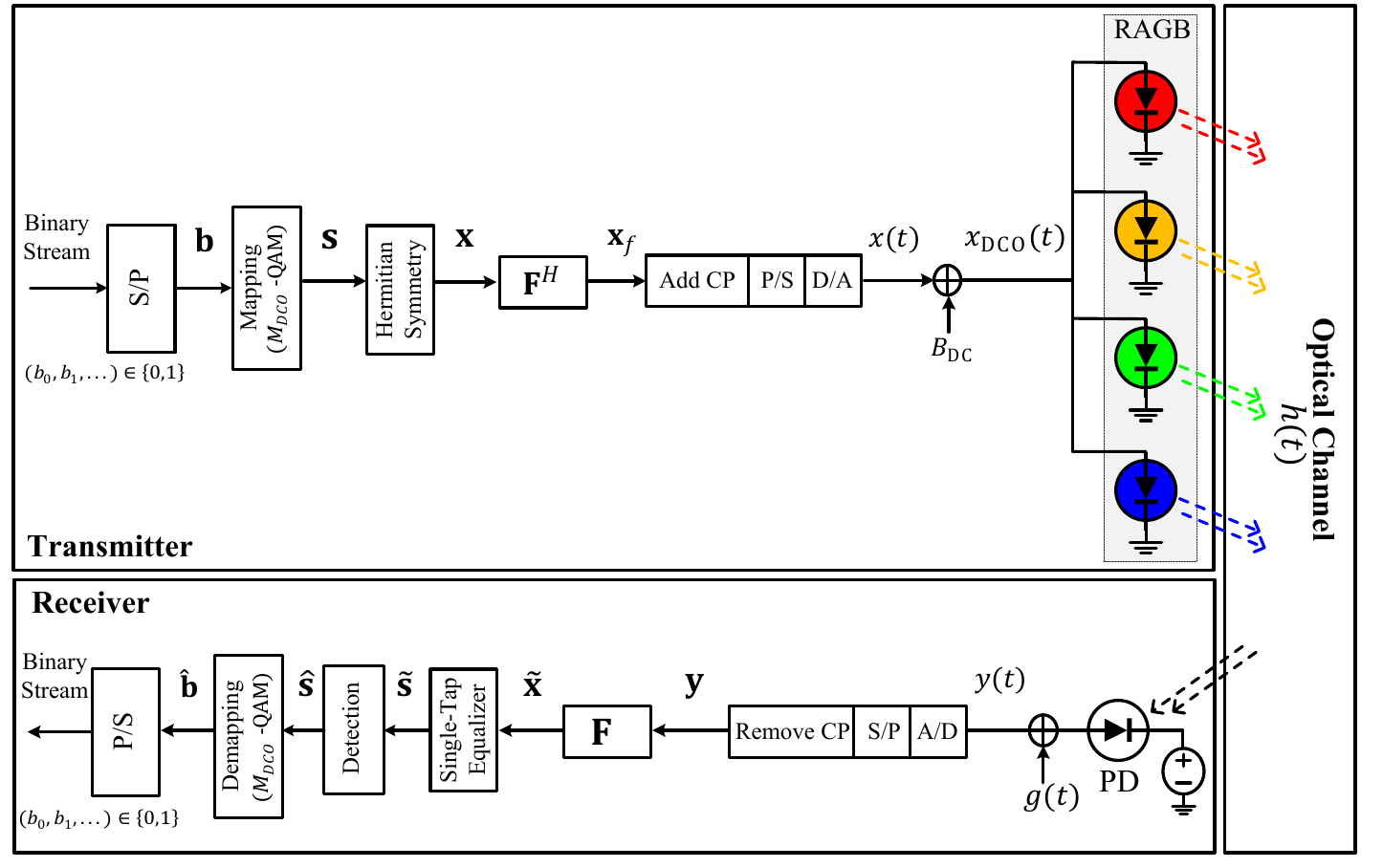}
  \caption{Conventional DCO‑OFDM with RAGB}
  \label{f:conventional}
\end{figure}

In a conventional DCO-OFDM transmitter driving a RAGB LED array, as depicted in Fig.~\ref{f:conventional}, the incoming binary stream
$(b_0,b_1,\ldots)\in\{0,1\}$ is grouped into frames of $(N/2-1)m_\mathrm{DCO}$ bits, where $N$ is the IDFT/IFFT size and $m_\mathrm{DCO}=\log_{2}M_\mathrm{DCO}$ is the number
of bits per $M_\mathrm{DCO}$‑QAM symbol. Each frame
\begin{equation}
\mathbf{b}=\bigl[b_0,\, b_1,\,\ldots,\, b_{{(N/2-1)m_\mathrm{DCO}}-1}\bigr]^{\mathrm T}\in\{0,1\}^{(N/2-1)m_\mathrm{DCO}}
\end{equation}
is parsed into $(N/2-1)$ words of $m_\mathrm{DCO}$ bits, and an $M_\mathrm{DCO}$‑ary QAM mapper converts these words into the complex vector
\begin{equation}
\mathbf{s}=[s_0, s_1,\ldots,s_{N/2-2}]^{\mathrm T}\in\mathbb C^{(N/2-1)}.
\end{equation}

Let $\mathbf{x}=[x_0,x_1,\ldots,x_{N-1}]^{\mathrm T}\in\mathbb{C}^{N}$ denote the frequency-domain OFDM symbol, where $x_k$ is the
DFT-bin value at index $k\in\{0,1,\ldots,N-1\}$. For even $N$, define the positive-frequency data-bearing index set $\mathcal{K}_{\mathrm{data}} \triangleq \{1,2,\ldots,N/2-1\}$. The QAM symbols are placed on the positive-frequency data subcarriers as $x_k=s_{k-1}$ for $k\in\mathcal{K}_{\mathrm{data}}$.
To obtain a real-valued time-domain waveform, $\mathbf{x}$ is constructed with Hermitian symmetry as
\begin{equation}
\mathbf{x}=
\bigl[\,0,\;\mathbf{s}^{\mathrm T},\;0,\;\operatorname{flip}(\mathbf{s}^{*})^{\mathrm T}\bigr]^{\mathrm T}
\in\mathbb{C}^{N},
\label{eq:HermitianX}
\end{equation}
where $\operatorname{flip}(\cdot)$ reverses the order of the vector elements. Subcarriers $k=0$ (DC) and $k=N/2$ (Nyquist) are nulled,
and the remaining negative-frequency bins satisfy $x_{N-k}=x_k^{*}$ for all $k\in\mathcal{K}_{\mathrm{data}}$.

Let $\mathbf{F}\in\mathbb{C}^{N\times N}$ denote the unitary discrete Fourier transform (DFT) matrix (implemented via fast Fourier transform, FFT) with
elements
\begin{equation}
[\mathbf{F}]_{k,n}
=\frac{1}{\sqrt{N}}\exp\!\Bigl(-\jmath\tfrac{2\pi kn}{N}\Bigr),
\qquad 0\le k,n\le N-1,
\label{eq:FFT_matrix}
\end{equation}
where $\jmath\triangleq\sqrt{-1}$. The time‑domain sample vector is obtained by the inverse DFT (IDFT, via IFFT)
\begin{equation}
\mathbf{x}_{f}=\mathbf{F}^{\mathrm H}\mathbf{x}\in\mathbb{R}^{N},
\label{eq:ifft_output}
\end{equation}
where $(\cdot)^{\mathrm H}$ denotes conjugate transpose. Because \(\mathbf{F}\) is unitary, the operation in~\eqref{eq:ifft_output} preserves signal energy, and the conjugate‑symmetric spectral pairs in \(\mathbf{x}\) make the imaginary parts of every sinusoidal component cancel, ensuring that $\mathbf{x}_{f}$ is strictly real. 

A cyclic prefix (CP) of length at least the channel memory is appended to $\mathbf{x}_{f}$, and the resulting block is serialized and converted to a continuous‑time electrical waveform $x(t)$ by a digital‑to‑analogue (D/A) converter, prior to RAGB LEDs driving for optical transmission. Since the intensity‑modulation/direct‑detection (IM/DD) front‑end can only process non‑negative drive levels, a constant DC bias $B_{\mathrm{DC}}$ is superimposed on the bipolar electrical signal $x(t)$
\begin{equation}
  x_{\mathrm{B}}(t)=x(t)+B_{\mathrm{DC}},
  \label{eq:dc_bias_add}
\end{equation}
and any remaining negative excursions are clipped,
\begin{equation}
  x_{\mathrm{DCO}}(t)=\max\!\bigl\{x_{\mathrm{B}}(t),\,0\bigr\},
  \label{eq:dc_bias_clip}
\end{equation}
yielding the unipolar drive waveform $x_{\mathrm{DCO}}(t)$ applied identically to each color channel of the RAGB array.

The bias level is parameterized as
\begin{equation}
  B_{\mathrm{DC}} = \mu\,\sigma_{x},
  \qquad
  \sigma_{x} \triangleq
  \sqrt{\mathbb{E}\!\bigl[(x(t)-\mathbb{E}[x(t)])^{2}\bigr]},
  \label{eq:dc_bias_mu}
\end{equation}
where $\sigma_x$ is the standard deviation of $x(t)$ and $\mu>0$ is the \emph{normalized bias factor}. For zero‑mean OFDM waveforms,
$\sigma_x$ coincides with the RMS amplitude.

In practice, $\mu$ can be estimated offline from $K$ statistically independent OFDM blocks
$\{\mathbf{x}_{f}^{(\kappa)}\}_{\kappa=1}^{K}$, where $\mathbf{x}_{f}^{(\kappa)}\in\mathbb{R}^{N}$. Denoting the samples by
$x_{f,n}^{(\kappa)}$ ($0\le n\le N-1$), the sample mean and standard deviation are
\begin{align}
  \bar{x}&=\frac{1}{KN}\sum_{\kappa=1}^{K}\sum_{n=0}^{N-1}x_{f,n}^{(\kappa)},
  \label{eq:x-hat}\\[1ex]
  \hat{\sigma}_{x} &=
  \sqrt{\frac{1}{KN}\sum_{\kappa=1}^{K}\sum_{n=0}^{N-1}
        \bigl(x_{f,n}^{(\kappa)}-\bar{x}\bigr)^{2}}.
  \label{eq:sigma-hat}
\end{align}
For each block, a preliminary bias factor
\begin{equation}
  \mu^{(\kappa)} =
  -\frac{\displaystyle\min_{0\le n<N}x_{f,n}^{(\kappa)}}{\hat{\sigma}_{x}}
  \label{eq:mu-k}
\end{equation}
is computed; averaging over $K$ blocks,
\begin{equation}
  \mu=\frac{1}{K}\sum_{\kappa=1}^{K}\mu^{(\kappa)},
  \label{eq:mu_mean}
\end{equation}
gives a bias value as the negative of the expected minimum sample,
$B_{\mathrm{DC}}\approx -\mathbb{E}\!\left[\min_{n}x_{f,n}^{(\kappa)}\right]$.
The ratio between the average electrical power of the biased waveform and the variance of $x(t)$ is
\begin{equation}
  B_{\mathrm{DC,dB}}\,[\mathrm{dB}]
  = 10\log_{10}\!\bigl(1+\mu^{2}\bigr),
  \label{eq:dc_bias_dB}
\end{equation}
which quantifies the power increase due to DC biasing \cite{6415964}.

At the receiver, a PD converts the incident optical power into an electrical current that is amplified, low‑pass filtered, sampled, and digitized. After CP removal, the $N$‑sample real vector $\mathbf{y}\in\mathbb{R}^{N}$ is obtained. Assuming CP length
$N_{\mathrm{CP}}\ge \vartheta-1$ and adopting the standard circular convolution model, the received block satisfies
\begin{equation}
\mathbf{y}=\mathbf{C}\mathbf{x}_{f}+\mathbf{g},
\qquad
\mathbf{g}\sim\mathcal{N}(\mathbf{0},\sigma_{g}^{2}\mathbf{I}_{N}),
\end{equation}
where $\mathbf{C}\in\mathbb{R}^{N\times N}$ is the circulant
channel matrix generated from the channel impulse response (CIR)
$\mathbf{h}=[h_{0}, h_{1},\ldots,h_{\vartheta-1}]^{\mathrm T}\in\mathbb{R}^{\vartheta}$ with $\vartheta$ denoting
the number of channel taps.

Applying the $N$‑point FFT gives
\begin{equation}
    \tilde{\mathbf{x}}= \mathbf{F}\mathbf{y}.
\end{equation}
 The subcarriers decouple and admit single-tap equalization (see Appendix~\ref{app:ofdm_single_tap}). Accordingly, for each data symbol index
$p=0,1,\ldots,\tfrac{N}{2}-2$ (corresponding to subcarrier $k=p+1\in\mathcal{K}_{\mathrm{data}}$), a zero-forcing (ZF) estimate of the
transmitted QAM symbol is
\begin{equation}
  \tilde{s}_{p} = \frac{\tilde{x}_{p+1}}{\Lambda_{p+1}},
  \qquad p = 0, 1, \ldots, \tfrac{N}{2}-2,
  \label{eq:s_hat}
\end{equation}
where $\Lambda_k$ is the $k$th frequency-domain channel gain defined in Appendix~\ref{app:ofdm_single_tap}. Maximum-likelihood detection
(MLD) is then performed on each equalized symbol,
\begin{equation}
  \hat{s}_{p}
  =\arg\min_{a_i\in\mathcal{M}_\mathrm{DCO}}
  \bigl|\tilde{s}_{p}-a_i\bigr|^{2},
  \qquad p = 0, 1, \ldots, \tfrac{N}{2}-2,
  \label{eq:ML_detection}
\end{equation}
where $\mathcal{M}_\mathrm{DCO}=\{a_0,\dots,a_{M_\mathrm{DCO}-1}\}$ is the $M_\mathrm{DCO}$-ary QAM constellation. The detected sequence $\{\hat{s}_{p}\}$ is demapped to
recover the transmitted bit stream.

\subsection{Proposed DC-Biased QCT Method}

\begin{figure}[t!]
	\centering
	\includegraphics[width=0.95\linewidth]{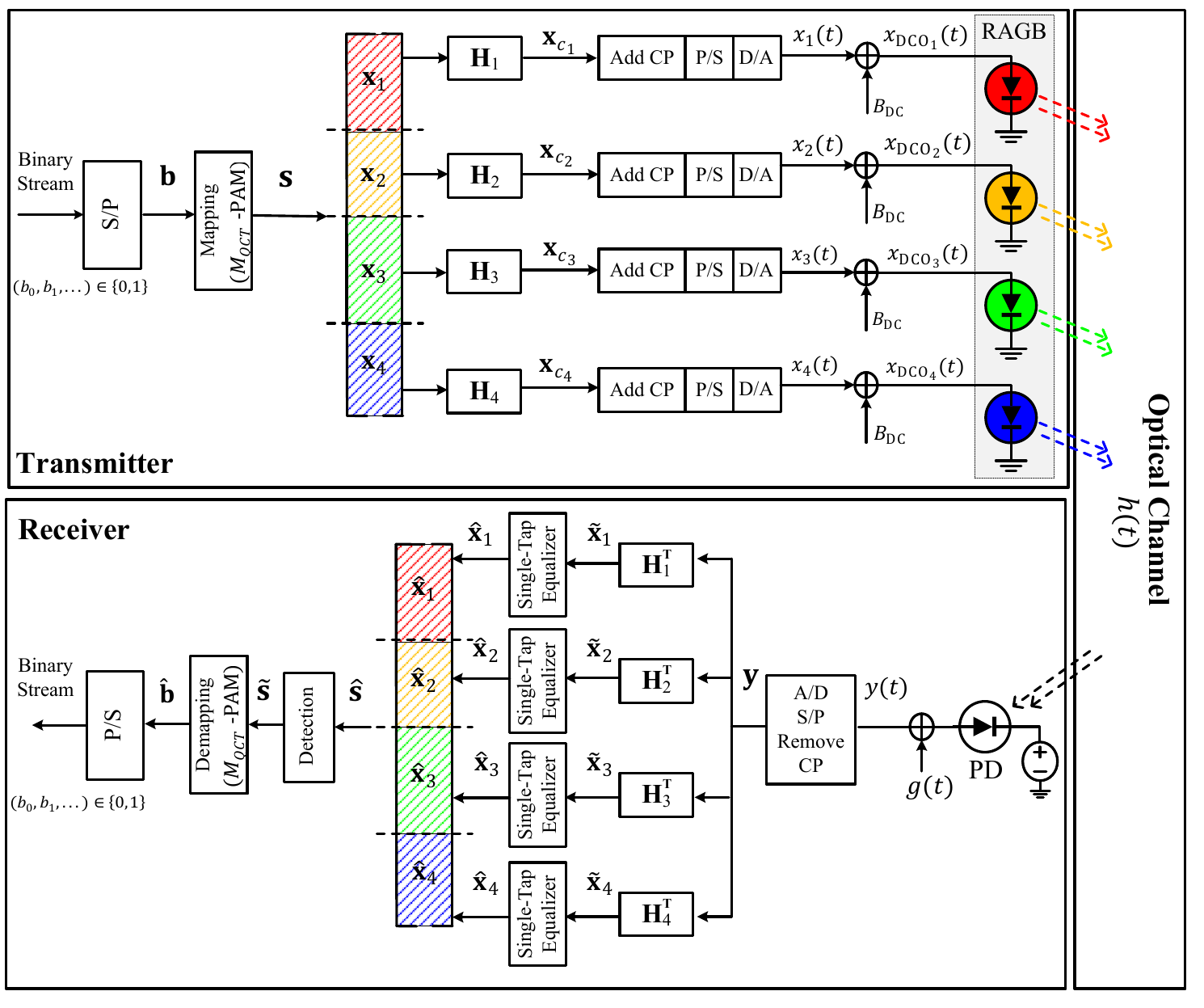}
	\caption{ Proposed DC-Biased QCT Method with RAGB}
	\label{f:proposedblock}
\end{figure}

The proposed QCT approach, illustrated in Fig.~\ref{f:proposedblock}, offers a filterless alternative by employing four‑channel separation
entirely in the digital domain rather than optical filtering. Specifically, four mutually orthogonal QCT synthesis matrices
\(
\mathbf{H}_{\nu}\in\mathbb{R}^{\,N\times N_0},\;
\nu\in\{1,2,3,4\},
\)
first introduced in~\cite{Cinemre2024QCTP}, operate on four disjoint subsets of $M_\mathrm{PAM}$‑PAM symbols, each subset containing $N_0\triangleq N/4$
symbols. By assigning one subset to each matrix, the system enables four parallel data streams transmitted via four independent LED
sources. The latter can be implemented either (i) by partitioning a multi‑chip white LED, modelled as a point source~\cite{7138560,7915761,6924005}, into four sub‑emitters, or (ii) by combining four distinct color LEDs (e.g., RAGB) that jointly generate white
light. In both realizations, the system supports the simultaneous transmission of four decorrelated optical channels without physical optical filters; in this work, the quadrichromatic RAGB implementation using four distinct color LEDs is adopted, motivated by the typically higher modulation bandwidth of emitting color chips compared with phosphor-converted white LEDs.

In this scheme, the incoming binary stream is sliced into successive frames of
\begin{equation}
\mathbf{b}=\bigl[b_0,\, b_1,\,\ldots,\, b_{Nm_\mathrm{PAM}-1}\bigr]^{\mathrm T}\in\{0,1\}^{Nm_\mathrm{PAM}}
\end{equation}
each containing $Nm_\mathrm{PAM}$ bits with $m_\mathrm{PAM}=\log_{2}M_\mathrm{PAM}$ bits per $M_\mathrm{PAM}$‑PAM symbol. Mapping delivers the real‑valued symbol vector
\begin{equation}
\mathbf{s}=[s_0,\ldots,s_{N-1}]^{\mathrm T}\in\mathbb R^{N},
\end{equation}
which is then partitioned into four equal‑length segments,
\begin{equation}
    \mathbf{s}
    =\bigl[\mathbf{x}_{1}^{\mathrm T}\!,
          \mathbf{x}_{2}^{\mathrm T}\!,
          \mathbf{x}_{3}^{\mathrm T}\!,
          \mathbf{x}_{4}^{\mathrm T}\bigr]^{\mathrm T}.
\end{equation}
The blocks $\mathbf{x}_{\nu}\in\mathbb R^{N_0}$, $\nu\in\{1,2,3,4\}$, drive the $\{\mathrm{R},\mathrm{A},\mathrm{G},\mathrm{B}\}$ LED branches,
respectively. The QCT synthesis matrices $\mathbf{H}_{\nu}\in\mathbb{R}^{N\times N_0}$ are applied to these symbol segments as
\begin{equation}\label{eq: x_c_nu}
  \mathbf{x}_{c,\nu} \;=\; \mathbf{H}_{\nu}\,\mathbf{x}_{\nu}.
\end{equation}
The resulting vectors $\mathbf{x}_{c,\nu}\in \mathbb{R}^{N}$ are each appended
with a CP and subsequently converted to analog signals for individual transmission through RAGB LEDs. Next, 
each branch
is DC‑biased according to \eqref{eq:dc_bias_mu}
\begin{equation}
  x_{B,\nu}(t)=x_{c,\nu}(t)+B_{\mathrm{DC}},
\end{equation}
and clipped at zero:
\begin{equation}
x_{\mathrm{DCO},\nu}(t)=\max\{x_{B,\nu}(t),\,0\}.
\end{equation}
The clipped waveforms propagate through the optical channel $h(t)$ and are captured by a single filterless PD:
\begin{equation}
  y(t)=h(t)\circledast
        \textstyle\sum_{\nu=1}^{4}x_{\mathrm{DCO},\nu}(t)+g(t),
\end{equation}
where $\circledast$ denotes linear convolution and $g(t)$ is AWGN. After A/D conversion, CP removal, and S/P conversion, the
$N$‑sample vector $\mathbf{y}\in\mathbb{R}^{N}$ is obtained. With $N_{\mathrm{CP}}> \vartheta-1$, the circular convolution model applies:
\begin{equation}
\mathbf{y}=\mathbf{C}\Bigl(\textstyle\sum_{\nu=1}^{4}\mathbf{x}_{c,\nu}\Bigr)+\mathbf{g},
\qquad \mathbf{g}\sim\mathcal{N}(\mathbf{0},\sigma_g^2\mathbf{I}_N),   
\end{equation}
where $\mathbf{C}\in\mathbb{R}^{N\times N}$ is the circulant channel matrix built from the $\vartheta$‑tap CIR
$\mathbf{h}=[h_0,\ldots,h_{\vartheta-1}]^{\mathrm T}\in\mathbb{R}^{\vartheta}$. Since the VLC channel is real-valued, the matched filter reduces to $\mathbf{C}^{\mathrm T}$, and the receiver forms
\begin{equation}
  \mathbf{z}\triangleq \mathbf{C}^{\mathrm T}\mathbf{y}
  = \mathbf{C}^{\mathrm T}\mathbf{C}\Bigl(\textstyle\sum_{\nu=1}^{4}\mathbf{x}_{c,\nu}\Bigr)+\mathbf{g}_z,
  \qquad \mathbf{g}_z\triangleq \mathbf{C}^{\mathrm T}\mathbf{g}.
  \label{eq:y_after_channel}
\end{equation}

Finally, the four data streams are separated by multiplying $\mathbf{z}\in\mathbb{R}^{N}$ with the transpose of the corresponding QCT
matrices, $\mathbf{H}^{\mathrm T}_\nu \in \mathbb{R}^{N_0\times N}$, yielding the recovered symbol estimates $\widetilde{\mathbf{x}}_\nu\in \mathbb{R}^{N_0}$:
\begin{equation}\label{eq: x_tilde_nu}
  \widetilde{\mathbf{x}}_{\nu}
  \;=\;
  \mathbf{H}_{\nu}^{\mathrm T}\,\mathbf{z}
  = \boldsymbol{\Lambda}_{\nu}\mathbf x_\nu + \mathbf H_\nu^{\mathrm T}\mathbf{g}_z, \qquad \nu\in\{1,2,3,4\}.
\end{equation}
By construction of the QCT matrices, the effective per‑stream channel matrices are diagonal:
\begin{equation}
  \boldsymbol{\Lambda}_{\nu}=
  \mathbf{H}_{\nu}^{\mathrm T}\mathbf C^{\mathrm T}\mathbf C\mathbf H_{\nu}
  =\operatorname{diag}(\Lambda_{\nu,0}, \Lambda_{\nu,1},\ldots,\Lambda_{\nu,N_0-1}),
  \label{eq:Lambdaqct}
\end{equation}
and the inter‑stream coupling terms vanish:
\begin{equation}
\mathbf{H}_{\nu}^{\mathrm T}\mathbf C^{\mathrm T}\mathbf
  C\mathbf H_{\nu'}=\mathbf 0,\qquad \nu'\neq\nu.
  \label{eq:Lambdaqctzero}
\end{equation}
A proof of \eqref{eq:Lambdaqct}-\eqref{eq:Lambdaqctzero} and the resulting single‑tap equalization is provided in
Appendix~\ref{app:qct_single_tap}.
Hence, for $\Lambda_{\nu,p}\neq 0$,
\begin{equation}
  \hat{x}_{\nu,p}=
  \frac{\widetilde{x}_{\nu,p}}{\Lambda_{\nu,p}},
  \qquad \forall\nu, \quad p=0, 1, \ldots,N_0-1
  \label{eq:x_hat_nu}
\end{equation}
The four equalized blocks are concatenated,
\begin{equation}
  \hat{\mathbf s}=
  \bigl[
    \hat{\mathbf x}_{1}^{\mathrm T},
    \hat{\mathbf x}_{2}^{\mathrm T},
    \hat{\mathbf x}_{3}^{\mathrm T},
    \hat{\mathbf x}_{4}^{\mathrm T}
  \bigr]^{\mathrm T},
\end{equation}
and passed through MLD (as in~\eqref{eq:ML_detection}, with $\mathcal{M}_\mathrm{DCO}$ interpreted as the $M_\mathrm{PAM}$‑PAM) followed by $M_\mathrm{PAM}$‑PAM
demapping to recover the original bit stream.

\section{Key Illumination Quality Metrics for VLC}\label{sec:illumination_metrics}

VLC luminaires must satisfy illumination and color quality constraints while simultaneously supporting data transmission. This section summarises three widely used lighting metrics; illuminance, CCT, and the CRI, and states how they are computed from the source SPD. Also, the simulation scenario and parameterization used in this work are described in Sec.~\ref{subsec:sim_params_NR}.

\subsection{Luminous Flux, Illuminance, and Uniformity}
\label{subsec:lux}

Let $\Phi_{e,\lambda}(\lambda)$ denote the source spectral radiant flux (radiant power per unit wavelength, in $\mathrm{W\,nm^{-1}}$) for $\lambda\in[380,780]~\mathrm{nm}$ (corresponding to the conventional photopic range \cite{ghassemlooy2017visible}), with the total radiant flux is $\Phi_e=\int_{380}^{780}\Phi_{e,\lambda}(\lambda)\,\mathrm{d}\lambda$ (in $\mathrm{W}$); for notational brevity, the SPD is denoted by $\Phi_e(\lambda)\triangleq \Phi_{e,\lambda}(\lambda)$ throughout the paper.

The \emph{luminous flux} $\Phi_{v}$ (in lumens, $\mathrm{lm}$) is obtained by weighting the SPD $\Phi_{e}(\lambda)$
with the CIE~1931 photopic luminosity function $\overline{V}(\lambda)$:
\begin{equation}
  \Phi_{v}
    = \mathcal{K}_{m}\!\int_{380}^{780}
        \Phi_{e}(\lambda)\,
        \overline{V}(\lambda)\,
        \mathrm{d}\lambda,
  \label{eq:luminous_flux}
\end{equation}
where $\mathcal{K}_{m}=683~\mathrm{lm\,W^{-1}}$ is the maximum luminous efficacy constant at $\lambda=555~\mathrm{nm}$.

\paragraph*{LoS point illuminance.}
Consider a Lambertian emitter $q$ with order $m_q$, LoS distance $d_{q}(r)$ to point $r$ on the
working plane, irradiance angle $\phi_q(r)$, and incidence angle $\psi_q(r)$.
The Lambertian model is parameterized by the semi-angle at half power $\xi_{1/2,q}$, which yields
\begin{equation}
m_q = -\frac{\ln 2}{\ln\!\bigl(\cos \Phi_{1/2,q}\bigr)}.
\label{eq:lambertian_order}
\end{equation}
The geometric irradiance factor is
\begin{equation}
    \zeta_{q\rightarrow r}
= \frac{m_q+1}{2\pi d_q^2(r)}\,
  \cos^{m_q}\!\bigl(\phi_q(r)\bigr)\,
  \cos\!\bigl(\psi_q(r)\bigr)\,
  \mathbbm{1}_{\{\psi_q(r)\le \Psi_{\mathrm{FOV}}\}}.
  \label{eq:geom_factor}
\end{equation}
and $\zeta_{q\rightarrow r}=0$ otherwise.
The LoS spectral irradiance (radiant power incident per unit area and per unit wavelength, in $\mathrm{W\,m^{-2}\,nm^{-1}}$) at $r$ is then
\begin{equation}
  E^{\mathrm{LoS}}_{e,\lambda}(r,\lambda)
  = \sum_{q=1}^{N_{\mathrm{tx}}} \Phi_{e,q}(\lambda)\, \zeta_q(r),
  \label{eq:E_LoS_spec}
\end{equation}
and the corresponding illuminance is
\begin{equation}
  E^{\mathrm{LoS}}(r)
  = \mathcal{K}_m \int_{380}^{780} E^{\mathrm{LoS}}_{e,\lambda}(r,\lambda)\, \overline{V}(\lambda)\, d\lambda.
  \label{eq:E_LoS_lux}
\end{equation}

\paragraph*{NLoS point illuminance (first-order reflections).}
In indoor environments, diffuse reflections from walls, ceiling, and floor contribute to illuminance and often improve
uniformity. Let $\mathcal{S}$ denote the set of reflecting surfaces, and let $\rho_s(\lambda)\in[0,1]$ be the
(typically weakly wavelength-dependent) diffuse reflectance of surface $s\in\mathcal{S}$.
For a differential surface element $dA$ located at $a\in A_s$,
the incident spectral irradiance is
\begin{equation}
  E^{\mathrm{inc}}_{e,\lambda}(a,\lambda)
  = \sum_{q=1}^{N_{\mathrm{tx}}} \Phi_{e,\lambda,q}(\lambda)\, \zeta_{q\rightarrow a},
  \label{eq:E_inc_surface}
\end{equation}
where $\zeta_{q\rightarrow a}$ has the same form as \eqref{eq:geom_factor} without an FOV constraint for the surface. Under Lambertian reflection, the reflected radiant exitance is $\rho_s(\lambda)\,E^{\mathrm{inc}}_{e,\lambda}(a,\lambda)$,
and the contribution of $dA$ to the spectral irradiance at $r$ is
\begin{equation}
  dE^{\mathrm{NLoS}}_{e,\lambda}(r,\lambda)
  =
  \rho_s(\lambda)\,E^{\mathrm{inc}}_{e,\lambda}(a,\lambda)\,
  \zeta_{a\rightarrow r}^{(1)}\; dA,
  \label{eq:dE_NLoS}
\end{equation}
with the Lambertian geometric factor, $m_a = 1$,
\begin{equation}
  \zeta_{a\rightarrow r}^{(1)}
  =
  \frac{1}{\pi\, d^2_{a}(r)}\,
  \cos\!\bigl(\phi_{a}(r)\bigr)\,
  \cos\!\bigl(\psi_{a}(r)\bigr)\,
  \mathbbm{1}_{\{\psi_{a}(r)\le \Psi_{\mathrm{FOV}}\}}.
  \label{eq:g_patch_to_r}
\end{equation}
where $d_{a}(r)$ is the distance from $a$ to $r$, $\phi_{a}(r)$ is the
angle between normal and the ray from $a$ to $r$, and $\psi_{a}(r)$ is the
incidence angle at the working plane.

Integrating over all reflecting surfaces yields the first-order NLoS spectral irradiance
\begin{equation}
  E^{\mathrm{NLoS}}_{e,\lambda}(r,\lambda)
  = \sum_{s\in\mathcal{S}} \int_{A_s}
      \rho_s(\lambda)\,E^{\mathrm{inc}}_{e,\lambda}(a,\lambda)\,
      \zeta_{a\rightarrow r}^{(1)}\, dA,
  \label{eq:E_NLoS_spec}
\end{equation}
and the NLoS illuminance
\begin{equation}
  E^{\mathrm{NLoS}}(r)
  = \mathcal{K}_m \int_{380}^{780} E^{\mathrm{NLoS}}_{e,\lambda}(r,\lambda)\, \overline{V}(\lambda)\, d\lambda.
  \label{eq:E_NLoS_lux}
\end{equation}
The total illuminance at $r$ is thus
\begin{equation}
  E(r) = E^{\mathrm{LoS}}(r) + E^{\mathrm{NLoS}}(r).
  \label{eq:E_total}
\end{equation}
Higher-order reflections can be incorporated by iterative evaluation, but in this work, \eqref{eq:E_total} is considered with only first-order reflections for simplicity.

\paragraph*{Illuminance uniformity.}
Let $\{r_k\}_{k=1}^{K}$ denote the set of sampled grid points on the working plane, and $E(r_k)$ be the
corresponding illuminance values computed via \eqref{eq:E_total}. The grid-average illuminance and minimum illuminance are
\begin{equation}
  \overline{E} = \frac{1}{K}\sum_{k=1}^{K} E(r_k),
  \qquad
  E_{\min} =q \min_{1\le k\le K} E(r_k),
  \label{eq:Eavg_Emin_grid}
\end{equation}
and the illuminance uniformity is defined as
\begin{equation}
  U_0 \triangleq \frac{E_{\min}}{\overline{E}}.
  \label{eq:uniformity}
\end{equation}

\subsection{Correlated Color Temperature (CCT)}
\label{subsec:CCT}

The perceived chromaticity of a white-light source is commonly quantified by its CCT, denoted $T_{\mathrm c}$ (in Kelvin). Given the source SPD $\Phi_e(\lambda)$, the
CIE~1931 tristimulus values are computed as
\begin{equation}
  (X,Y,Z)=
  \int_{380}^{780}
    \Phi_{e}(\lambda)\,
    \bigl(\overline{x}(\lambda),\overline{y}(\lambda),\overline{z}(\lambda)\bigr)\,
    \mathrm{d}\lambda,
  \label{eq:XYZ}
\end{equation}
where $\overline{x}(\lambda)$, $\overline{y}(\lambda)$, and $\overline{z}(\lambda)$ are the CIE~1931 color matching
functions \cite{commission1931commission, smith1931cie}. Any positive scaling of $\Phi_e(\lambda)$ scales $(X,Y,Z)$ equally and therefore does not affect $(x,y)$ or the computed CCT. The normalized chromaticity coordinates are then
\begin{equation}
  x = \frac{X}{X+Y+Z}, \qquad
  y = \frac{Y}{X+Y+Z}.
  \label{eq:xy_coords}
\end{equation}

To determine $T_{\mathrm c}$, $(x,y)$ is mapped to the CIE\,1960 UCS $(u,v)$ space; $T_{\mathrm c}$ is then the temperature of the black‑body radiator on the Planckian locus whose $(u,v)$ chromaticity is closest (in Euclidean distance) to that of the test source. The proximity is typically evaluated via Robertson’s interpolation scheme or a lookup table of $(u,v,T)$ values~\cite{macadam1937projective,Carter2018CIE0C}.
\begin{figure*}[t]
 \centering
  \includegraphics[width=0.95\linewidth]{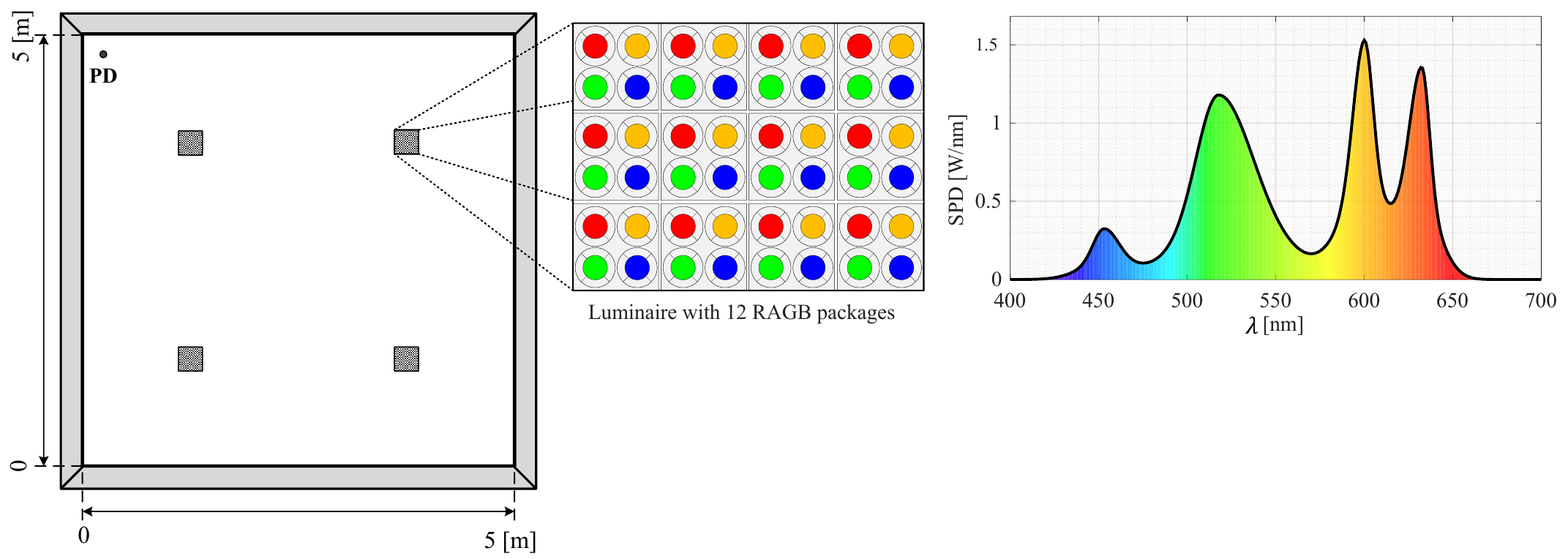}
  \caption{Top view of the simulated room with four ceiling-mounted luminaires, each containing 12 LUXEON-C RAGB LED packages. The inset shows the luminaire SPD $\Phi_e(\lambda)$ as a function of wavelength, illustrating the RAGB spectral characteristics.}
  \label{fig:locpsd}
\end{figure*}

\subsection{Color Rendering Index (CRI)}
\label{subsec:CRI}

The general CRI $R_{\mathrm a}\in[0,100]$ quantifies how faithfully a test light source renders
object colors relative to a reference illuminant having the same correlated color temperature $T_{\mathrm c}$.
By definition, $R_{\mathrm a}=100$ corresponds to perfect color rendition with respect to the reference.

Let $\Phi_e^{t}(\lambda)$ and $\Phi_e^{r}(\lambda)$ denote the SPDs of the test and reference illuminants, respectively, and
let $R_i(\lambda)$ denote the reflectance spectrum of the $i$th CIE test-color sample (TCS01-TCS08), $i=1,\dots,8$.
The corresponding CIE~1931 tristimulus values are computed as
\begin{equation}
  (X_i^{(\cdot)},Y_i^{(\cdot)},Z_i^{(\cdot)})
  =
  \int_{380}^{780}
    \Phi_{e}^{(\cdot)}(\lambda)\, R_i(\lambda)\,
    \bigl(\overline{x}(\lambda),\overline{y}(\lambda),\overline{z}(\lambda)\bigr)\,
    \mathrm{d}\lambda,
  \label{eq:XYZ_TCS}
\end{equation}
where $(\cdot)\in\{t,r\}$ denotes test or reference.

The resulting $(X_i^t,Y_i^t,Z_i^t)$ and $(X_i^r,Y_i^r,Z_i^r)$ are transformed to CIE~1960 UCS chromaticity coordinates
$(u_i^t,v_i^t)$ and $(u_i^r,v_i^r)$. A von~Kries chromatic-adaptation transform is then applied to align the test-source
chromaticity to that of the reference \cite{takasaki1969kries}. The adapted coordinates are subsequently mapped to the
CIE~1964 $U^{*}V^{*}W^{*}$ space, yielding $(U_i^t,V_i^t,W_i^t)$ and $(U_i^r,V_i^r,W_i^r)$ for the test and reference,
respectively \cite{wyszecki1963proposal}. The color difference for sample $i$ is defined as
\begin{equation}
  \Delta \mathcal{E}_i
  = \sqrt{(U^t_{i} - U^r_{i})^{2}
          + (V^t_{i} - V^r_{i})^{2}
          + (W^t_{i} - W^r_{i})^{2}}.
  \label{eq:deltaE}
\end{equation}
The corresponding special color rendering indices are
\begin{equation}
  \mathcal{R}_i = 100 - 4.6\,\Delta \mathcal{E}_i, \qquad i=1,\dots,8,
  \label{eq:Ri}
\end{equation}
and the general CRI is defined as their arithmetic mean:
\begin{equation}
  \mathcal{R}_{\mathrm a} = \frac{1}{8}\sum_{i=1}^{8} \mathcal{R}_i.
  \label{eq:CRI}
\end{equation}

\section{Numerical Results}
\subsection{Simulation Parameters}
\label{subsec:sim_params_NR}

The numerical results are obtained for the indoor VLC scenario illustrated in Fig.~\ref{fig:locpsd}; the main geometric,
device, and noise parameters are summarized in Table~\ref{tab:sim_NR}. The room dimensions are $5\times 5\times 3~\mathrm{m}^3$ and
four ceiling-mounted luminaires are deployed at the locations listed in Table~\ref{tab:sim_NR}. Each luminaire contains
$12$ LUXEON-C RAGB LED packages (i.e., $48$ packages in total), where the $\{\mathrm{R},\mathrm{A},\mathrm{G},\mathrm{B}\}$ emitters within
a package are co-located and oriented normal to the ceiling. A fixed total electrical power budget of $P_{\mathrm{tot}}=480~\mathrm{W}$ is enforced and kept identical across all considered modulation
schemes. To maintain a common white-light operating point (fixed chromaticity and color quality) throughout the simulations, the
color mixture ratio (CMR) \cite{9329045} of the four color channels 
$\mathrm{R}:\mathrm{A}:\mathrm{G}:\mathrm{B}=70:70:140:20$ (normalized weights $0.2333:0.2333:0.4667:0.0667$) \cite{luxeon_datasheet},
and the same mixture vector is used for all simulation runs to ensure that performance differences are not confounded by changes in the operating color point.

\begin{table}[t]
  \centering
\caption{\(H\)-model parameters for LUXEON-C RAGB channels \cite{8546751} (\(\lambda_{p,c}\): peak wavelength; \(\Delta\lambda_{1,c}\), \(\Delta\lambda_{2,c}\): lower-/upper-side spectral width parameters; \(k_{1,c}\): relative amplitude factor; \(k_{2,c}\): width-shaping factor).}
  \label{tab:Hparams_NR}
  \begin{tabular}{@{}lccccc@{}}
    \toprule
    Channel
    & \(\lambda_{p,c}\) (\(\mathrm{nm}\))
    & \(\Delta\lambda_{1,c}\) (\(\mathrm{nm}\))
    & \(\Delta\lambda_{2,c}\) (\(\mathrm{nm}\))
    & \(k_{1,c}\)
    & \(k_{2,c}\) \\
    \midrule
    Red   & 632.5 & 23.84 & 14.74 & 2 & 6 \\
    Amber & 600.0 & 19.66 & 14.97 & 2 & 5 \\
    Green & 517.7 & 29.38 & 45.21 & 2 & 3 \\
    Blue  & 453.0 & 18.99 & 25.50 & 2 & 5 \\
    \bottomrule
  \end{tabular}
\end{table}

\paragraph*{Per-channel SPD.}
Following Sec.~\ref{subsec:lux}, to model the emitted spectra, the SPD of each color channel \(c\in\{\mathrm R,\mathrm A,\mathrm G,\mathrm B\}\) is written as
\begin{equation}
  \Phi_{e,c}(\lambda) \;=\; \tau_{c}\,\mathcal{S}_{c}(\lambda),
  \label{eq:abs_spd}
\end{equation}
where \(\mathcal{S}_{c}(\lambda)\) is the dimensionless \(H\)-model shape \cite{he2010model}
\begin{equation}
\mathcal{S}_{c}(\lambda)=
\frac{\exp\!\Bigl(-\dfrac{(\lambda-\lambda_{p,c})^{2}}{\Delta\lambda_{c}^{2}}\Bigr)
      +k_{1,c}\,\exp\!\Bigl(-k_{2,c}\dfrac{(\lambda-\lambda_{p,c})^{2}}{\Delta\lambda_{c}^{2}}\Bigr)}
     {1+k_{1,c}},
\label{eq:SPDc}
\end{equation}
with the asymmetric width
\begin{equation}
  \Delta\lambda_{c}
    = \begin{cases}
         \Delta\lambda_{1,c}, & \lambda < \lambda_{p,c},\\
         \Delta\lambda_{2,c}, & \lambda \ge \lambda_{p,c}.
       \end{cases}
  \label{eq:Delta_lambda_piecewise}
\end{equation}
The scale factor \(\tau_{c}>0\) sets the absolute radiant-flux level of channel \(c\) and is chosen to meet a desired
radiant-flux target \(\Phi_{e,c}^{\mathrm{tot}}\) (in \(\mathrm{W}\)):
\begin{equation}
  \tau_{c} \;=\; \frac{\Phi_{e,c}^{\mathrm{tot}}}{S_c},
  \qquad
  S_c \triangleq \int_{380}^{780} \mathcal{S}_{c}(\lambda)\,\mathrm{d}\lambda.
  \label{eq:Ac_radiant}
\end{equation}
The composite SPD of an RAGB package is then
\begin{equation}
  \Phi_{e}(\lambda)=\sum_{c\in\{\mathrm R,\mathrm A,\mathrm G,\mathrm B\}}\Phi_{e,c}(\lambda).
  \label{eq:SPD_compose}
\end{equation}
The resulting composite SPD (for a single luminaire) based on parameter values listed in Table~\ref{tab:Hparams_NR} is shown in
Fig.~\ref{fig:locpsd}.
For the CSK benchmark, the receiver employs ideal rectangular color filters with unit gain and pass-bands
\([612,680]\), \([575,612]\), \([483,575]\), and \([400,483]\)~\(\mathrm{nm}\) for \(\{\mathrm R,\mathrm A,\mathrm G,\mathrm B\}\),
respectively.
Additive noise is modelled as AWGN with two-sided current power spectral density
\(10^{-22}~\mathrm{A^{2}/Hz}\).
For analytical clarity, the channel DC gain is normalized to unity.

\begin{table}[t]
  \centering
  \caption{Simulation parameters used in the numerical evaluation.}
  \label{tab:sim_NR}

  {%
  \setlength{\tabcolsep}{4pt}
  \renewcommand{\arraystretch}{1.15}
  \scriptsize
  \colorlet{secrow}{gray!15}

  \begin{tabular}{@{}p{0.50\linewidth}p{0.46\linewidth}@{}}
    \toprule
    \textbf{Parameter} & \textbf{Value} \\
\midrule
    \rowcolor{secrow}\multicolumn{2}{@{}l@{}}{\textbf{Room geometry and luminaire deployment}}\\
    \midrule
    Room dimensions  & $5\times 5\times 3~\mathrm{m}$ \\
    Number of luminaires & $4$ \\
    RAGB packages per luminaire & $12$ \\
    Luminaire center coordinates ($x,y,z$) &
    \begin{tabular}[t]{@{}l@{}}
      $(1.25,\,3.75,\,3)~\mathrm{m}$ \\
      $(3.75,\,3.75,\,3)~\mathrm{m}$ \\
      $(1.25,\,1.25,\,3)~\mathrm{m}$ \\
      $(3.75,\,1.25,\,3)~\mathrm{m}$
    \end{tabular} \\
    RAGB package pointing direction & Elevation: $-90^\circ$, Azimuth: $0^\circ$ \\
\midrule
    \rowcolor{secrow}\multicolumn{2}{@{}l@{}}{\textbf{Transmitter (LED) characteristics}}\\
    \midrule
    RAGB package model & LUXEON-C Color RAGB \\
    Electrical power per RAGB package & $10~\mathrm{W}$ \\
    Semi-angle at half power($\xi_{1/2,q}$) & $60^\circ$ \\
\midrule
    \rowcolor{secrow}\multicolumn{2}{@{}l@{}}{\textbf{Receiver (PD) configuration}}\\
    \midrule
    PD coordinate ($x,y,z$) & $(0.25,\,4.75,\,0.85)~\mathrm{m}$ \\
    PD pointing direction & Elevation: $90^\circ$, Azimuth: $0^\circ$ \\
    PD field of view ($\Psi_{\mathrm{FOV}}$) & $90^\circ$ \\
    PD physical area & $7\times 10^{-6}~\mathrm{m}^2$ \\

\midrule
    \rowcolor{secrow}\multicolumn{2}{@{}l@{}}{\textbf{Propagation and reflection model}}\\
    \midrule
    Wall reflectance &
    \begin{tabular}[t]{@{}l@{}}
      Wavelength-dependent $\rho(\lambda)$; \\
      measured data from \cite{5682214}, \\
      avg.\ $\bar{\rho}\approx 0.72$
    \end{tabular} \\
    Channel DC gain (normalized) & $1$ \\

    \midrule
    \rowcolor{secrow}\multicolumn{2}{@{}l@{}}{\textbf{System and noise settings}}\\
    \midrule
    System bandwidth & $30~\mathrm{MHz}$ \\
    Total electrical power $P_{\mathrm{tot}}$ & $480~\mathrm{W}$ \\
    AWGN PSD & $1.0\times 10^{-22}~\mathrm{A^{2}/Hz}$ \\
\midrule
    \rowcolor{secrow}\multicolumn{2}{@{}l@{}}{\textbf{Spectral filters and source metrics}}\\
    \midrule
    Optical filter passbands $\{\mathrm{R},\mathrm{A},\mathrm{G},\mathrm{B}\}$&
    \begin{tabular}[t]{@{}l@{}}
      Lower: $\{612,\,575,\,483,\,400\}~\mathrm{nm}$ \\
      Upper: $\{680,\,612,\,575,\,483\}~\mathrm{nm}$
    \end{tabular} \\
    Filter gains (CSK only) & $1$ (ideal) \\
    Source metrics & $T_{\mathrm c}\approx 3858~\mathrm{K}$, $R_{\mathrm a}\approx 81.5$ \\

    \bottomrule
  \end{tabular}
  } 
\end{table}

\subsection{Fair comparison protocol}
\label{sec:fairness}

Three RAGB VLC transceiver configurations: (i) RAGB CSK with a four filter and four PD receiver, (ii) conventional RAGB DCO-OFDM, and (iii) the proposed filterless RAGB QCT scheme, are examined under a common physical deployment in terms of \emph{communication-layer} comparisons (BER, distortion, PAPR) and \emph{illumination-layer} outcomes (illuminance, uniformity, CCT/CRI, flicker).

\paragraph*{Common channel model under a point-source approximation.}
The discrete-time CIRs from each luminaire to the considered PD location are obtained via the ray-tracing procedure in~\cite{5682214}, including the LoS component and the NLoS components due to first-order reflections. Following the standard point source abstraction used in indoor VLC channel modeling~\cite{7138560}, each luminaire is represented by an equivalent point source located at its geometric center; thus, each luminaire is characterized by a common real-valued $\vartheta$-tap discrete-time CIR $\mathbf{h}$ and, after CP removal, the corresponding circulant channel matrix $\mathbf{C}\in\mathbb{R}^{N\times N}$. This approximation is primarily governed by the luminaire aperture relative to the Tx-Rx distance and is known to yield negligible deviations in optical path loss and transmission bandwidth for moderate array sizes while maintaining acceptable delay-spread accuracy; similar simplifications are routinely adopted in MIMO-VLC simulations~\cite{7564676}. Since the $\{\mathrm R,\mathrm A,\mathrm G,\mathrm B\}$ emitters within a package are co-located, the same baseband CIR is applied to all four color branches, i.e., $h_{\mathrm{R}}(t)=h_{\mathrm{A}}(t)=h_{\mathrm{G}}(t)=h_{\mathrm{B}}(t)\triangleq h(t)$, which is the key propagation assumption enabling QCT stream orthogonalization. Finally, consistent with our ray-tracing results and~\cite{7823364,5682214,10640091}, after normalizing the CIR such that $\sum_{\ell=0}^{\vartheta-1} h_{\ell}=1$, the channel is LoS dominated: the first tap $h_{0}$ typically captures $70$-$85\%$ of the received power, the dominant postcursor taps (e.g., $h_{1}$ and $h_{2}$) contribute approximately $15$-$25\%$ and $5$-$10\%$, respectively, and the remaining taps carry only a minor residual fraction. In the numerical evaluation, the QCT synthesis matrices $\{\mathbf{H}_{\nu}\}_{\nu=1}^{4}$ are designed once using a single reference channel measurement $\mathbf{h}^{(0)}$ (equivalently, $\mathbf{C}^{(0)}$) and then kept fixed, whereas the reported results are obtained by averaging over $N_{\mathrm{MC}}$ independent channel realizations $\{\mathbf{h}^{(i)}\}_{i=1}^{N_{\mathrm{MC}}}$ drawn randomly from the same LoS-dominated tap profile.

\paragraph*{Optical-power constraint and SNR normalization.}
The comparison is performed under a strict mean optical-power constraint. For any real electrical waveform $x(t)$, the emitted IM/DD waveform is
$x_{\mathrm{DCO}}(t)=\max\{x(t)+B_{\mathrm{DC}},0\}$ and $P_{\mathrm{opt}}\triangleq\mathbb{E}[x_{\mathrm{DCO}}(t)]$. The total mean optical power
(summed over the same set of active LED packages) is held constant across schemes; for QCT, the per-branch bias factors are chosen such that the
per-branch mean optical power matches that of DCO-OFDM, i.e., $P_{\mathrm{opt}}^{\mathrm{QCT}}=P_{\mathrm{opt}}^{\mathrm{DCO}}$. To avoid bias-dependent noise rescaling, $\sigma_g^2$ is set using an unbiased, unclipped reference waveform at each SNR point
and then kept fixed, so that any performance change reflects interference and clipping behavior.

\paragraph*{Illumination and flicker constraints.}
In addition to the communication metrics, all schemes are evaluated under common lighting constraints.
For a fixed room geometry and a fixed luminaire SPD, the photometric and
colorimetric quantities depend on the time-average emitted optical power and spectrum, rather than
on the specific modulation format. Accordingly, under the matched mean optical-power constraint imposed
above (with the same set of active LED packages), the illuminance on the working plane is computed from
the common SPD using \eqref{eq:E_total}, and the corresponding uniformity,
$U_{0}$ in \eqref{eq:uniformity}, is used to assess compliance with typical
office-lighting recommendations (e.g., $\overline{E}\in[300,1500]~\mathrm{lx}$ and $U_{0}\ge 0.7$) \cite{5089970}. In
parallel, color quality is characterized via the correlated color temperature $T_{\mathrm c}$ obtained from
\eqref{eq:XYZ}--\eqref{eq:xy_coords} and the general color rendering index $R_{\mathrm a}$ in \eqref{eq:CRI};
for interior workplaces, common targets are $R_{\mathrm a}\ge 80$ and $T_{\mathrm c}\approx 3500~\mathrm{K}$
\cite{10.3390/electronics11030346,10.1021/cm100010z}. Finally, to ensure that the transmitted optical
waveform is free of perceptible temporal light modulation (flicker) \cite{6852085}, flicker is quantified
on the nonnegative emitted optical waveform $y(t)\ge 0$ (after DC biasing and clipping) using the percent
flicker and flicker index:
\begin{equation}
  \mathrm{PF} = 100\,\frac{y_{\max}-y_{\min}}{y_{\max}+y_{\min}} \quad [\%],
  \label{eq:percent_flicker}
\end{equation}
where $y_{\max}=\max_t\,y(t)$ and $y_{\min}=\min_t\,y(t)$, and
\begin{equation}
  \mathrm{FI} = \frac{\sum_{t:\,y(t)>\bar{y}}\!\bigl(y(t)-\bar{y}\bigr)}{\sum_{t} y(t)},
  \label{eq:flicker_index}
\end{equation}
where $\bar{y}=\mathbb{E}[y(t)]$ is the time-average optical power (in discrete-time simulations, the
sums are taken over the sampled time indices). The numerical illuminance maps, uniformity, and color/flicker
metrics are reported in Sec.~\ref{subsec:perf_results}.

\paragraph*{Spectral efficiency matching.}
The net spectral efficiency (including CP overhead where applicable) is
\begin{align}
  \eta_{\mathrm{DCO}} &= \frac{\bigl(\frac{N}{2}-1\bigr)\log_2(M_{\mathrm{QAM}})}{N+N_{\mathrm{CP}}}\quad \mathrm{bits/s/Hz},\\
  \eta_{\mathrm{QCT}} &= \frac{N\log_2(M_{\mathrm{PAM}})}{N+N_{\mathrm{CP}}}\quad \mathrm{bits/s/Hz},\\
  \eta_{\mathrm{CSK}} &= \log_2(M_{\mathrm{CSK}})\quad \mathrm{bits/s/Hz},
\end{align}
where $M_{\mathrm{QAM}}$ is the QAM order for DCO-OFDM, $M_{\mathrm{PAM}}$ is the PAM order per QCT branch, and $M_{\mathrm{CSK}}$ is the CSK
constellation size. The modulation orders are selected to equalize net throughput,
i.e., $\eta_{\mathrm{DCO}}=\eta_{\mathrm{QCT}}=\eta_{\mathrm{CSK}}$; in particular, matching $\eta_{\mathrm{QCT}}=\eta_{\mathrm{DCO}}$ implies
$M_{\mathrm{PAM}}\approx \sqrt{M_{\mathrm{QAM}}}$ for large $N$.

\paragraph*{PAPR analysis.}
In this paper, PAPR is evaluated on the bipolar time-domain blocks prior to DC biasing and clipping, since it
directly reflects the dynamic-range requirement of the electrical waveform generated by the IDFT/QCT synthesis.
Accordingly, in DCO-OFDM, the real-valued IDFT output  $\mathbf{x}_{f}\in\mathbb{R}^{N}$ in \eqref{eq:ifft_output} is used, whereas
in QCT, the per-branch synthesized blocks $\mathbf{x}_{c,\nu}=\mathbf{H}_{\nu}\mathbf{x}_{\nu}\in\mathbb{R}^{N}$ in
\eqref{eq: x_c_nu} is utilized. For any real length-$N$ block $\mathbf{u}=[u_0,\ldots,u_{N-1}]^{\mathsf T}$, define
\begin{equation}
\mathrm{PAPR}(\mathbf{u})
\triangleq
10\log_{10}\!\left(
\frac{\max_{0\le n\le N-1}|u_n|^{2}}
{\frac{1}{N}\sum_{n=0}^{N-1}|u_n|^{2}}
\right)\ \text{dB}.
\label{eq:papr_def_systemmodel}
\end{equation}
Hence, the DCO-OFDM block PAPR is
$\mathrm{PAPR}_{\mathrm{DCO}}\triangleq \mathrm{PAPR}(\mathbf{x}_{f})$.
For QCT, the per-branch PAPRs
$\mathrm{PAPR}_{\nu}\triangleq \mathrm{PAPR}(\mathbf{x}_{c,\nu})$ are computed and report the \emph{worst-branch PAPR per block} as
\begin{equation}
\mathrm{PAPR}_{\mathrm{QCT}}
\triangleq
\max_{\nu\in\{1,2,3,4\}}\mathrm{PAPR}(\mathbf{x}_{c,\nu}).
\label{eq:papr_qct_worst_systemmodel}
\end{equation}
The PAPR complementary CDF (CCDF) is then defined for either scheme as
\begin{equation}
\mathrm{CCDF}(\gamma)\triangleq \Pr\{\mathrm{PAPR}>\gamma\},
\label{eq:papr_ccdf_systemmodel}
\end{equation}
where $\gamma$ is a PAPR threshold (in dB); in Monte Carlo simulations, $\mathrm{CCDF}(\gamma)$ is estimated by the fraction of
transmitted blocks whose measured PAPR exceeds $\gamma$.

\subsection{Performance Results and Discussion}
\label{subsec:perf_results}

\begin{figure*}
    \centering
    \includegraphics[width=1\linewidth]{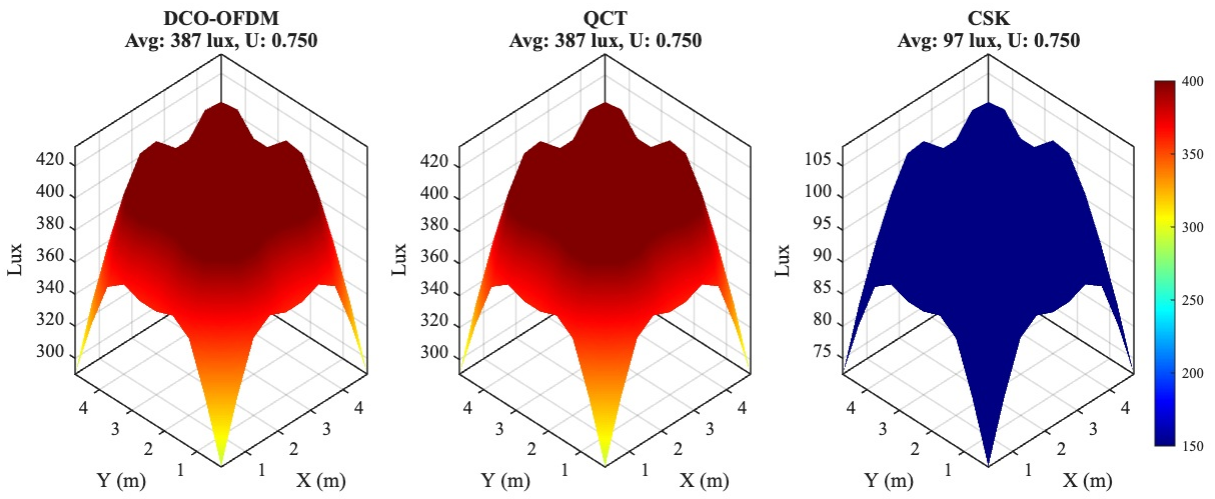}
    \caption{Spatial distribution of illuminance at $z = 0.85$ m for LoS + NLoS.}
    \label{fig:LoS+NLoS Aydınlatma}
\end{figure*}

\paragraph*{Illumination.}
Because all schemes use the same RAGB spectral model and the same average CMR, the spectral shape of the time-averaged emission is common across the three schemes. Consequently, colour-quality metrics are essentially modulation invariant for a fixed environment: in our setup the received light yields $\mathrm{CRI}\approx 81.46$ and $\mathrm{CCT}\approx 3858$~K under LoS-only propagation and $\mathrm{CCT}\approx 3829$~K under LoS+NLoS, satisfying the typical indoor targets summarized in Sec.~\ref{sec:fairness}.

In contrast, the illuminance level scales with the time-average emitted optical power under the fixed per-package electrical power budget in Table~\ref{tab:sim_NR}. Under this physically relevant constraint, the OFDM-based schemes (DCO-OFDM and QCT) achieve substantially larger mean emitted optical power than CSK, which translates into a much higher luminous flux and illuminance. Specifically, the total luminous flux (summed over the active LED packages) is
$\Phi_{\mathrm{tot}}\!=\!14398.6~\mathrm{lm}$ for both DCO-OFDM and QCT, whereas $\Phi_{\mathrm{tot}}\!=\!3600.0~\mathrm{lm}$ for CSK.
Accordingly, CSK provides the lowest average illuminance ($72.6~\mathrm{lx}$ for LoS and $96.6~\mathrm{lx}$ for LoS+NLoS), while both OFDM-based schemes provide substantially higher and nearly identical illuminance levels ($\sim\!291~\mathrm{lx}$ for LoS and $\sim\!387~\mathrm{lx}$ for LoS+NLoS), i.e., an approximately four-fold increase over CSK, as it can be seen in Fig.~\ref{fig:LoS+NLoS Aydınlatma}.
Incorporating first-order reflections increases the mean illuminance and improves spatial uniformity (from $U_{0}=0.5521$ (LoS) to $U_{0}=0.7505$ (LoS+NLoS)) while reducing the coefficient of variation (from $0.206$ to $0.096$), in line with the office-lighting and uniformity criteria reviewed in Sec.~\ref{sec:fairness}. 

Since illuminance scales approximately linearly with the number of LED packages, achieving a $300~\mathrm{lx}$ target under LoS-only propagation would require approximately $199$ RAGB packages for CSK, compared to approximately $50$ packages for either QCT or DCO-OFDM (a $74.9\%$ reduction). Finally, temporal-light-modulation metrics (percent flicker and flicker index) as defined in \eqref{eq:percent_flicker} and \eqref{eq:flicker_index} are nonzero (e.g., $\mathrm{PF}\approx 100\%$ for $16$-QAM DCO-OFDM and $\approx 40\%$ for $4$-PAM QCT) when evaluated on the instantaneous baseband waveform, but the employed $30~\mathrm{MHz}$ signaling bandwidth is orders of magnitude above the human flicker-fusion range ($<200~\mathrm{Hz}$); hence, no perceptible flicker is expected and the IEEE~802.15.7 flicker constraints are satisfied \cite{6852085}.

\begin{figure*}[t]
    \centering
    \begin{subfigure}[t]{0.48\linewidth}
        \centering
        \includegraphics[width=\linewidth]{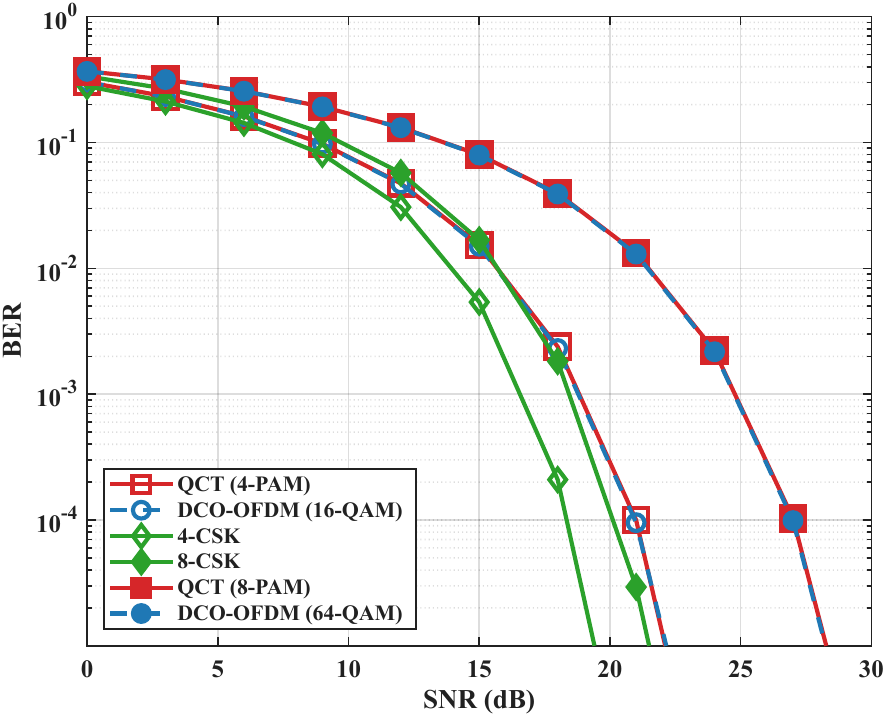}
        \caption{BER versus the normalized SNR (Sec.~\ref{sec:fairness}) for throughput-matched operating points:
        $\eta\approx 2~\mathrm{bits/s/Hz}$ ($M_\mathrm{CSK}=M_\mathrm{QCT} =4$, $M_\mathrm{DCO}=16$) and
        $\eta\approx 4~\mathrm{bits/s/Hz}$ ($M_\mathrm{CSK}=M_\mathrm{QCT} =8$, $M_\mathrm{DCO}=64$).}
        \label{fig:ber_comparison}
    \end{subfigure}
    \hfill
    \begin{subfigure}[t]{0.48\linewidth}
        \centering
        \includegraphics[width=\linewidth]{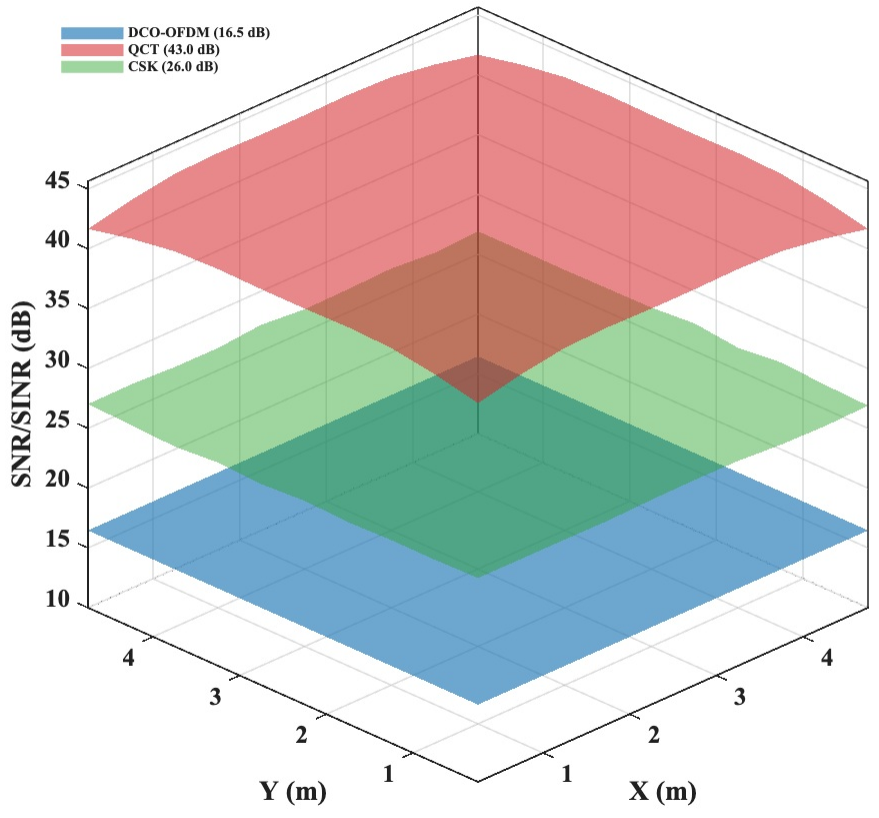}
        \caption{Spatial distribution of the effective SNR/SINR on the $z=0.85$~m working plane under LoS+NLoS propagation for
        $\eta\approx 2~\mathrm{bits/s/Hz}$.}
        \label{fig:16QAMLONLOSSSNR}
    \end{subfigure}

    \caption{BER performance and spatial link-quality comparison of DCO-OFDM, QCT, and CSK.}
    \label{fig:combined_results}
\end{figure*}

\paragraph*{BER and spatial SNR/SINR comparison.}
Fig.~\ref{fig:combined_results}\subref{fig:ber_comparison} presents the BER baseline versus the
normalized SNR definition adopted in Sec.~\ref{sec:fairness}. Under this linear reference model, the spectrally matched QCT-OFDM and DCO-OFDM configurations exhibit nearly identical BER, since
both reduce to orthogonal modulation with linear equalization in AWGN. Fig.~\ref{fig:combined_results}\subref{fig:16QAMLONLOSSSNR} complements this baseline by reporting the spatial distribution of the effective link quality on the $z=0.85$~m working plane under LoS+NLoS propagation when the practical IM/DD front-end is captured via an AWGN\,+\,clipping-noise model. The corresponding plane-averaged SNR/SINR values extracted from the spatial maps are reported in Table~\ref{tab:snr_summary}; for CSK, the metric is an SINR that additionally incorporates inter-channel crosstalk, whereas for the OFDM-based schemes it reduces to an SNR. Even if the AWGN benchmark BER curves of QCT-OFDM and DCO-OFDM overlap, QCT-OFDM attains a higher effective SNR over the receiver plane because its clipping-distortion power is smaller under the same mean optical-power constraint. Note that the absolute SNR/SINR levels are not directly comparable
across the two spectral-efficiency operating points, since the matched mean optical power per branch (and hence the bias level) changes (cf.\ Table~\ref{tab:perf512_summary}); therefore, comparisons should be made within each operating point rather than across operating points.

\begin{table}[t]
\centering
\caption{$z=0.85$~$\mathrm{m}$ working plane ($10\times 10$ uniform grid) effective SNR/SINR (dB) (including clipping distortion as in Sec.~IV-B)  under LoS and LoS+NLoS propagation for two spectral-efficiency operating points
(\(\eta\approx 2\) and \(\eta\approx 4\)~$\mathrm{bits/s/Hz}$).}
\label{tab:snr_summary}

{%
\setlength{\tabcolsep}{5.0pt}
\renewcommand{\arraystretch}{1.12}
\scriptsize
\begin{tabular}{@{}lcccc@{}}
\toprule
& \multicolumn{2}{c}{LoS} & \multicolumn{2}{c}{LoS+NLoS} \\
\cmidrule(lr){2-3}\cmidrule(lr){4-5}
\textbf{Scheme} & \(\eta\approx 2\) & \(\eta\approx 4\) & \(\eta\approx 2\) & \(\eta\approx 4\) \\
\midrule
DCO-OFDM & 16.55 & 31.21 & 16.55 & 31.23 \\
QCT & \textbf{40.94} & \textbf{46.79} & \textbf{42.94} & \textbf{48.95} \\
CSK       & 25.88 & 26.06 & 26.02 & 26.24 \\
\bottomrule
\end{tabular}
{\vspace{0.3em}
\begin{minipage}{0.98\columnwidth}
\end{minipage}}
}
\end{table}

\paragraph*{Clipping and EVM comparison.}
Table~\ref{tab:perf512_summary} compares DCO-OFDM and the proposed QCT-OFDM at $N=512$ under the fair-comparison protocol in Sec.~\ref{sec:fairness}. Consistent with the point-source luminaire abstraction, the $12$ RAGB packages within each luminaire are co-located and driven synchronously; hence, each luminaire is represented by four equivalent color emitters $\{\mathrm{R},\mathrm{A},\mathrm{G},\mathrm{B}\}$, and all metrics are reported per color branch.
As required by Sec.~\ref{sec:fairness}, the per-branch mean optical power is strictly matched, (i.e., $P_{\mathrm{opt}}^{\mathrm{QCT}}=P_{\mathrm{opt}}^{\mathrm{DCO}}$); the mapping between $P_{\mathrm{opt}}$, $\sigma_x$, and the normalized bias $\mu=B_{\mathrm{DC}}/\sigma_x$ follows directly from Appendix~\ref{app:gaussian_clipping_moments}. In addition to Monte Carlo simulation measurements, the corresponding normalized baseband  $P_{\mathrm{opt}}$, $P_{\mathrm{elec}}$, $P_{\mathrm{clip}}$ power moments are computed using the closed-form expressions in Appendix~\ref{app:gaussian_clipping_moments}; the resulting theoretical values closely track the simulated ones and satisfy the power identity in~\eqref{eq:power_identity_app}.

\begin{table*}[t]
\centering
\caption{Clipping, EVM, and power statistics at $N=512$ under matched mean optical power per equivalent color branch.
Theoretical $P_{\mathrm{elec}}$ and $P_{\mathrm{clip}}$ are computed from Appendix~\ref{app:gaussian_clipping_moments}
[\eqref{eq:EY2_app} and \eqref{eq:Ec2_app}], and the identity check corresponds to \eqref{eq:power_identity_app}.}
\label{tab:perf512_summary}

{%
\setlength{\tabcolsep}{2.2pt}
\renewcommand{\arraystretch}{1.10}
\setlength{\heavyrulewidth}{0.2em}
\setlength{\lightrulewidth}{0.1em}
\setlength{\cmidrulewidth}{0.020em}
\scriptsize
\colorlet{mshade}{gray!15}

\begin{tabular}{@{}c l c c c c c c c@{}}
\toprule
\textbf{$P_{\mathrm{opt}}$ per branch}
& \textbf{OFDM scheme (mod.)}
& \textbf{Bias}
& \textbf{Clipping}
& \textbf{$P_{\mathrm{elec}}$ per branch}
& \textbf{$P_{\mathrm{clip}}$ per branch}
& \textbf{EVM}
& \textbf{Identity check}
& \textbf{Rel.\ err} \\
\cmidrule(lr){1-2}\cmidrule(lr){3-8}

& 
& \cellcolor{mshade}$B_{\mathrm{DC,dB}}$
& \cellcolor{mshade}Clip rate [\%]
& \cellcolor{mshade}$P_{\mathrm{elec}}$ (sim.)
& \cellcolor{mshade}$P_{\mathrm{clip}}$ (sim.)
& \cellcolor{mshade}EVM$_{\mathrm{clip}}$ [\%]
& \cellcolor{mshade}$\mathbb{E}[Y^2]+\mathbb{E}[c^2]$
& \\

\cmidrule(lr){3-8}
&
& \cellcolor{mshade!30}$\mu$
& \cellcolor{mshade!30}ClipTX [dB]
& \cellcolor{mshade!30}$P_{\mathrm{elec}}$ (th.)
& \cellcolor{mshade!30}$P_{\mathrm{clip}}$ (th.)
& \cellcolor{mshade!30}EVM$_{\mathrm{tot}}$ [\%]
& \cellcolor{mshade!30}$\sigma_x^2+B_{\mathrm{DC}}^2$
& \\
\midrule

\multirow{6}{*}{6.348}
& \multirow{2}{*}{DCO-OFDM (16-QAM)}
& \cellcolor{mshade}$7.00$
& \cellcolor{mshade}$2.256$
& \cellcolor{mshade}$49.864$
& \cellcolor{mshade}$5.660\times 10^{-2}$
& \cellcolor{mshade}$7.46$
& \cellcolor{mshade}$49.9206$
& \multirow{2}{*}{$4.286\times 10^{-5}$} \\

\cmidrule(lr){3-8}
&
& \cellcolor{mshade!30}$2.003$
& \cellcolor{mshade!30}$-22.66$
& \cellcolor{mshade!30}$49.8615$
& \cellcolor{mshade!30}$5.696\times 10^{-2}$
& \cellcolor{mshade!30}$35.44$
& \cellcolor{mshade!30}$49.9184$
& \\

\cmidrule(lr){2-9}
& \multirow{2}{*}{QCT (4-PAM)}
& \cellcolor{mshade}$12.34$
& \cellcolor{mshade}$0.002$
& \cellcolor{mshade}$42.7969$
& \cellcolor{mshade}$5.740\times 10^{-6}$
& \cellcolor{mshade}$0.15$
& \cellcolor{mshade}$42.7969$
& \multirow{2}{*}{$8.231\times 10^{-6}$} \\

\cmidrule(lr){3-8}
&
& \cellcolor{mshade!30}$4.015$
& \cellcolor{mshade!30}$-56.67$
& \cellcolor{mshade!30}$42.7966$
& \cellcolor{mshade!30}$7.202\times 10^{-6}$
& \cellcolor{mshade!30}$33.06$
& \cellcolor{mshade!30}$42.7966$
& \\

\addlinespace[0.25em]
\midrule

\multirow{6}{*}{19.41}
& \multirow{2}{*}{DCO-OFDM (64-QAM)}
& \cellcolor{mshade}$10.00$
& \cellcolor{mshade}$0.134$
& \cellcolor{mshade}$418.381$
& \cellcolor{mshade}$8.333\times 10^{-3}$
& \cellcolor{mshade}$1.40$
& \cellcolor{mshade}$418.389$
& \multirow{2}{*}{$5.110\times 10^{-6}$} \\

\cmidrule(lr){3-8}
&
& \cellcolor{mshade!30}$3.000$
& \cellcolor{mshade!30}$-37.35$
& \cellcolor{mshade!30}$418.383$
& \cellcolor{mshade!30}$8.512\times 10^{-3}$
& \cellcolor{mshade!30}$34.27$
& \cellcolor{mshade!30}$418.391$
& \\

\cmidrule(lr){2-9}
& \multirow{2}{*}{QCT (8-PAM)}
& \cellcolor{mshade}$15.67$
& \cellcolor{mshade}$\approx 0$
& \cellcolor{mshade}$387.248$
& \cellcolor{mshade}$0$
& \cellcolor{mshade}$\approx 0$
& \cellcolor{mshade}$387.248$
& \multirow{2}{*}{$4.856\times 10^{-6}$} \\

\cmidrule(lr){3-8}
&
& \cellcolor{mshade!30}$5.991$
& \cellcolor{mshade!30}$\ll -300$
& \cellcolor{mshade!30}$387.246$
& \cellcolor{mshade!30}$5.398\times 10^{-10}$
& \cellcolor{mshade!30}$33.05$
& \cellcolor{mshade!30}$387.246$
& \\

\bottomrule
\end{tabular}
} 
\end{table*}

Table~\ref{tab:perf512_summary} shows that, at the same mean optical-power budget, QCT-OFDM yields orders-of-magnitude lower clipping than DCO-OFDM. At $P_{\mathrm{opt}}\approx 6.348$, the clipped-sample rate decreases from $2.256\%$ to $0.002\%$, and the clipping power drops from $5.66\times 10^{-2}$ to $5.74\times 10^{-6}$ per branch ($\approx 40$~dB reduction), which directly reduces the clipping-only EVM from $7.46\%$ to $0.15\%$. At $P_{\mathrm{opt}}\approx 19.41$, DCO-OFDM still exhibits nonzero clipping rate ($0.134\%$) and a finite clipping-only EVM ($1.40\%$), whereas QCT-OFDM clipping becomes negligible (with theoretical $P_{\mathrm{clip}}$ on the order of $10^{-10}$), yielding an essentially zero clipping only EVM. The improvement in the average EVM$_{\mathrm{tot}}$ is comparatively modest because it is averaged
over a wide SNR range in which low-to-mid SNR points are AWGN-limited; therefore, EVM$_{\mathrm{clip}}$ is the more diagnostic metric for isolating nonlinear distortion under a matched optical-power constraint.

The larger reported bias in dB for QCT does not contradict the matched mean optical-power condition because $B_{\mathrm{DC, dB}}$ is determined by the normalized bias $\mu$, not by the absolute offset $B_{\mathrm{DC}}$ alone.
Under matched $P_{\mathrm{opt}}$, the absolute $B_{\mathrm{DC}}$ values are of the same order in both schemes, but the per-branch unbiased standard deviation $\sigma_x$ is smaller for QCT; consequently, QCT operates at a larger $\mu$ (and hence $B_{\mathrm{DC,dB}}$) and thus a much smaller negative-tail probability $\Pr\{X<-B_{\mathrm{DC}}\}=\bar{F}_Z(\mu)$ and clipping moment $\mathbb{E}[c^2]$ (Appendix~\ref{app:gaussian_clipping_moments}), explaining the near-zero clipping and the vanishing clipping-only EVM observed in Table~\ref{tab:perf512_summary}.

\begin{figure}[t]
    \centering
    \includegraphics[width=1\linewidth]{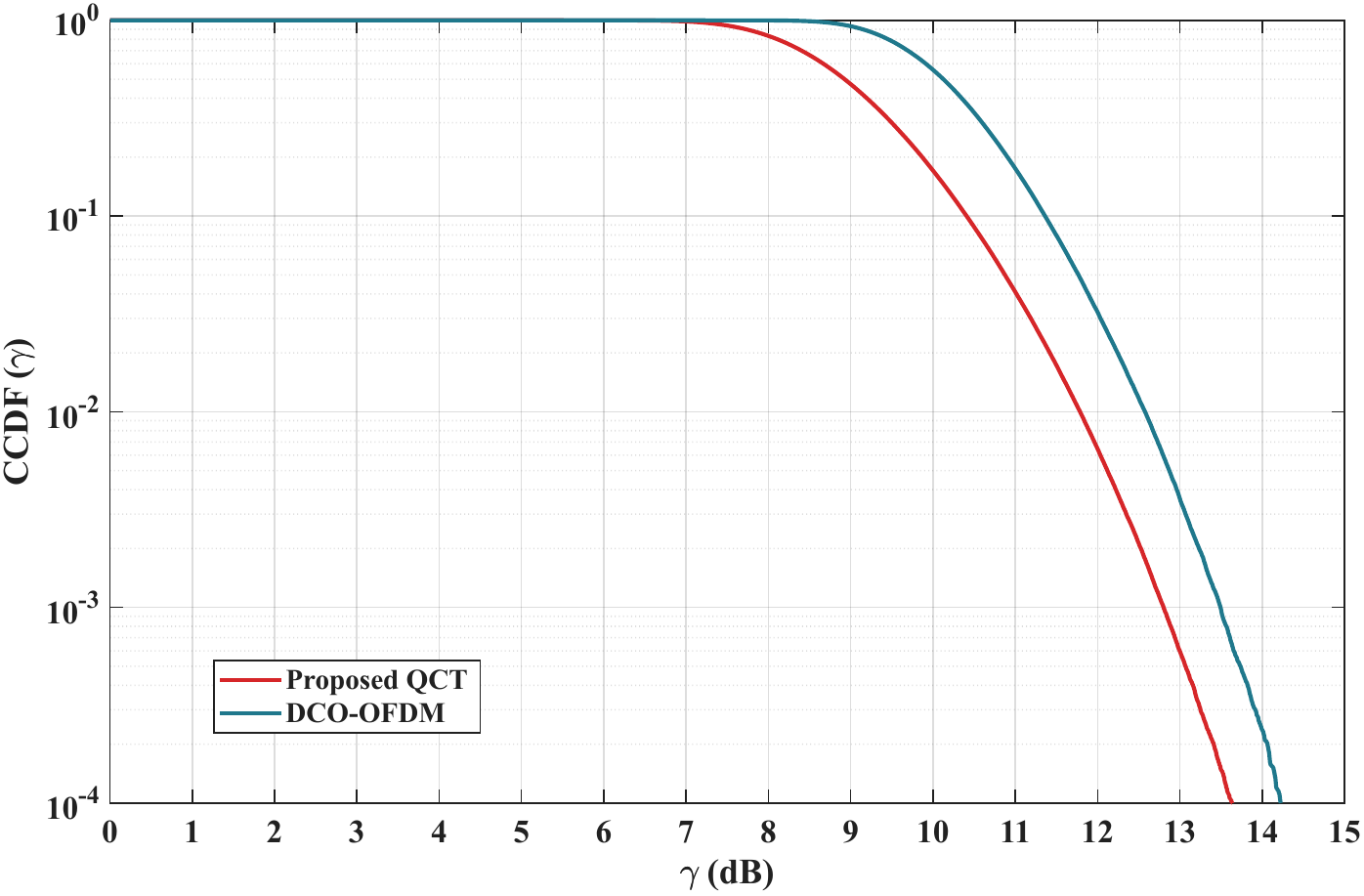} 
\caption{CCDF of the PAPR comparison}
    \label{fig:papr_comparison}
\end{figure}

\paragraph*{PAPR analysis.}
Fig.~\ref{fig:papr_comparison} depicts the CCDF of the pre-bias PAPR defined in
\eqref{eq:papr_def_systemmodel}-\eqref{eq:papr_ccdf_systemmodel}, where the proposed QCT-OFDM achieves a consistent $0.7$--$1$~dB PAPR reduction relative to DCO-OFDM over the full CCDF range; for example, at $\mathrm{CCDF}=10^{-3}$ the required threshold decreases from $\approx 13.5$~dB (DCO-OFDM) to $\approx 12.8$~dB (QCT-OFDM). This improvement is primarily attributable to the different waveform-synthesis structures: in DCO-OFDM, each time-domain sample of the IDFT output $\mathbf{x}_f$ in \eqref{eq:ifft_output} results from the superposition of $N/2-1$ independently modulated subcarriers, so occasional near-coherent additions can produce pronounced peaks; in contrast, each QCT branch waveform $\mathbf{x}_{c,\nu}=\mathbf{H}_\nu\mathbf{x}_\nu$ in \eqref{eq: x_c_nu} is synthesized from a lower-dimensional real symbol vector $\mathbf{x}_\nu\in\mathbb{R}^{N/4}$ via $\mathbf{H}_\nu\in\mathbb{R}^{N\times (N/4)}$, meaning that each sample depends on fewer independent data degrees of freedom per branch, which reduces the probability of extreme constructive superposition and tightens the upper tail of the PAPR distribution. Practically, the resulting PAPR reduction lowers the required electrical dynamic range (DAC) and provides additional margin against the DC-biasing/clipping tradeoff in IM/DD transmission, complementing the clipping-power and EVM$_{\mathrm{clip}}$ trends reported above.

\section{Conclusion}

This paper developed and evaluated a DC-biased QCT transmission scheme for RAGB VLC that shifts color separation and interference mitigation from the optical domain to low-complexity digital processing. By exploiting the commuting symmetry operators of the matched-filtered circulant VLC channel, four orthogonal QCT subspaces were constructed and diagonalized, enabling four interference-free streams to be recovered from a single filterless photodiode through per-stream single-tap equalization. A unified comparison against RAGB-CSK and DCO-OFDM was carried out under consistent throughput, channel, and illumination constraints.

Numerical results in a representative indoor room model showed that the proposed QCT approach significantly improves practical link quality in IM/DD operation: it achieves up to $48.95~\mathrm{dB}$ average effective SNR on the working plane, yielding $15.1-22.7~\mathrm{dB}$ gain over filter-based CSK and $15.6-26.4~\mathrm{dB}$ gain over DCO-OFDM under the adopted clipping noise model, while matching the BER performance of throughput-matched DCO-OFDM in the linear AWGN baseline. Under matched mean optical power, QCT substantially suppresses nonlinear distortion, exhibiting orders-of-magnitude lower clipping power and a modest but consistent PAPR reduction compared with DCO-OFDM. In addition, using the same luminaire SPD and deployment geometry, the achieved lighting quality remains within typical indoor targets (CCT and CRI) while providing substantially higher illuminance than the CSK benchmark.

Future work will focus on validating robustness to practical color branch mismatches (e.g., wavelength-dependent gains, LED bandwidth differences, and non-identical impulse responses across R/A/G/B), extending the design to adaptive/estimated channels and measured hardware nonlinearity, and experimentally demonstrating filterless QCT reception in a real multi-color luminaire prototype.

\appendices

\section{Single-Tap Equalization}
\label{app:single_tap}

\subsection{Reminder: Single-Tap Equalization in Optical OFDM}
\label{app:ofdm_single_tap}

Let $\mathbf{x}\in\mathbb{C}^{N}$ denote the vector of frequency-domain subcarrier symbols. The corresponding
time-domain OFDM block is generated via the unitary IDFT as in \eqref{eq:ifft_output},
\begin{equation}
  \mathbf{x}_{f}=\mathbf{F}^{\mathrm H}\mathbf{x},
  \label{eq:ifft_output_app}
\end{equation}
where $\mathbf{F}\in\mathbb{C}^{N\times N}$ is the unitary DFT matrix defined in \eqref{eq:FFT_matrix}.
When a real-valued time-domain waveform is required, $\mathbf{x}$ is constructed to satisfy Hermitian symmetry
so that $\mathbf{x}_{f}\in\mathbb{R}^{N}$.

A CP of length $N_{\mathrm{CP}}\ge \vartheta-1$ is appended at the transmitter and removed at the
receiver. Under standard OFDM assumptions, an LTI $\vartheta$-tap channel over the $N$-sample block and
proper CP alignment such that interference is avoided, the received vector
$\mathbf{y}\in\mathbb{R}^{N}$ follows the circular convolution model
\begin{equation}
\mathbf{y}=\mathbf{C}\mathbf{x}_{f}+\mathbf{g},
  \label{eq:ofdm_circ_model_app_new}
\end{equation}
where $\mathbf{g}\sim\mathcal{N}(\mathbf{0},\sigma_g^{2}\mathbf{I}_{N})$ and $\mathbf{C}\in\mathbb{R}^{N\times N}$
is the circulant channel matrix generated by the zero-padded impulse response
\begin{equation}
  \mathbf{h}^{\mathrm{zp}}\triangleq [h_{0},h_{1},\ldots,h_{\vartheta-1},0,\ldots,0]^{\mathrm T}\in\mathbb{R}^{N},
\end{equation}
i.e., the first column of $\mathbf{C}$ equals $\mathbf{h}^{\mathrm{zp}}$.

A fundamental property of circulant matrices is diagonalization by the DFT basis. In particular,
\begin{equation}
  \mathbf{F}\mathbf{C}\mathbf{F}^{\mathrm H}
  =\boldsymbol{\Lambda}
  =\operatorname{diag}\!\bigl(\sqrt{N}\,\mathbf{F}\mathbf{h}^{\mathrm{zp}}\bigr)
  =\operatorname{diag}(\Lambda_0,\ldots,\Lambda_{N-1}),
  \label{eq:circulant_dft_app}
\end{equation}
where $\boldsymbol{\Lambda}\in\mathbb{C}^{N\times N}$ is diagonal and $\Lambda_k$ is the $k$th frequency-domain channel
gain given by
\begin{equation}
  \Lambda_k \triangleq \sum_{\ell=0}^{\vartheta-1} h_{\ell}\exp\!\Bigl(-\jmath\tfrac{2\pi k\ell}{N}\Bigr),
  \qquad k=0,1,\ldots,N-1.
  \label{eq:Lambda_entries}
\end{equation}

Applying the $N$-point DFT to \eqref{eq:ofdm_circ_model_app_new} and using $\mathbf{x}_{f}=\mathbf{F}^{\mathrm H}\mathbf{x}$
yields
\begin{equation}
  \tilde{\mathbf{x}}
  \triangleq \mathbf{F}\mathbf{y}
  = \mathbf{F}\mathbf{C}\mathbf{x}_{f}+\mathbf{F}\mathbf{g}
  = \boldsymbol{\Lambda}\,\mathbf{x}+\tilde{\mathbf{g}},
  \qquad \tilde{\mathbf{g}}\triangleq \mathbf{F}\mathbf{g}.
  \label{eq:tilde_x_app_new}
\end{equation}
Since $\mathbf{F}$ is unitary, the noise remains white in the transform domain in the sense that
$\mathbb{E}[\tilde{\mathbf{g}}\tilde{\mathbf{g}}^{\mathrm H}]=\sigma_g^2\mathbf{I}_N$. Consequently, the subcarriers
decouple as
\begin{equation}
  \tilde{x}_k = \Lambda_k\,x_k + \tilde{g}_k,
  \qquad k=0,1,\ldots,N-1,
  \label{eq:per_tone_app_new}
\end{equation}
which enables single-tap equalization. For $\Lambda_k\neq 0$, the ZF equalizer is
\begin{equation}
  \widehat{x}_k=\frac{\tilde{x}_k}{\Lambda_k},\qquad k=0,1,\ldots,N-1.
  \label{eq:zf_one_tap_app_new}
\end{equation}
To mitigate noise enhancement, a standard regularized alternative is the minimum mean square error (MMSE) single-tap equalizer
\begin{equation}
  \widehat{x}_k
  =\frac{\Lambda_k^{*}}{|\Lambda_k|^2+\sigma_g^2/\sigma_x^2}\,\tilde{x}_k,
  \qquad k=0,1,\ldots,N-1,
  \label{eq:mmse_one_tap_app_new}
\end{equation}
where $\sigma_s^2 \triangleq \mathbb{E}[|x_k|^2]$ denotes the average symbol energy (equivalently, the variance for zero-mean constellations)
on the data-bearing subcarriers $k\in\mathcal{K}_{\mathrm{data}}$.

\subsection{Single-Tap Equalization in the Proposed QCT Receiver}
\label{app:qct_single_tap}

It is demonstrated that the proposed QCT construction yields (i) four interference-free streams and (ii) a single-tap equalizer for each stream.

\subsubsection{Matched-filtered circulant model}
Assume the CP condition holds, so the useful sample model is
\begin{equation}
  \mathbf{y}=\mathbf{C}\mathbf{x}_{\mathrm{sp}}+\mathbf{g},
  \label{eq:qct_rx_app}
\end{equation}
where $\mathbf{C}\in\mathbb{R}^{N\times N}$ is circulant, $\mathbf{g}\sim\mathcal{N}(\mathbf{0},\sigma_g^{2}\mathbf{I}_{N})$, and
$\mathbf{x}_{\mathrm{sp}}\in\mathbb{R}^{N}$ is the superposition of four QCT-coded streams
$\mathbf{x}_\nu\in\mathbb{R}^{N_0}$, $\nu\in\{1,2,3,4\}$:
\begin{equation}
  \mathbf{x}_{\mathrm{sp}}
  =\sum_{\nu=1}^{4}\mathbf{x}_{c,\nu}
  =\sum_{\nu=1}^{4}\mathbf{H}_\nu \mathbf{x}_\nu.
  \label{eq:qct_tx_sum_app}
\end{equation}
Here $\mathbf{H}_\nu\in\mathbb{R}^{N\times N_0}$ are the QCT matrices.
The receiver applies the matched filter $\mathbf{C}^{\mathrm T}$
and obtains
\begin{equation}
  \mathbf{z}= \mathbf{C}^{\mathrm T}\mathbf{y}
  = \underbrace{\mathbf{C}^{\mathrm T}\mathbf{C}}_{\mathbf{G}}\mathbf{x}_{\mathrm{sp}} + \mathbf{g}_z,
  \qquad \mathbf{g}_z= \mathbf{C}^{\mathrm T}\mathbf{g}.
  \label{eq:qct_mf_model_app}
\end{equation}
Here $\mathbf{G}=\mathbf{C}^{\mathrm T}\mathbf{C}$ is real, symmetric, and circulant, and $\mathbf{g}_z$ is generally colored with
$\mathbb{E}[\mathbf{g}_z\mathbf{g}_z^{\mathrm T}]=\sigma_g^2\mathbf{G}$.
Substituting \eqref{eq:qct_tx_sum_app} into
\eqref{eq:qct_mf_model_app} yields
\begin{equation}
  \mathbf{z}=\sum_{\nu=1}^{4}\mathbf{G}\mathbf{H}_\nu\mathbf{x}_\nu + \mathbf{g}_z.
  \label{eq:qct_mf_model2_app}
\end{equation}

Projecting $\mathbf{z}$ onto the $\nu$th stream gives
\begin{equation}
  \widetilde{\mathbf{x}}_{\nu}
  \triangleq \mathbf{H}_\nu^{\mathrm T}\mathbf{z}
  =
  \sum_{\nu'=1}^{4}\mathbf{H}_\nu^{\mathrm T}\mathbf{G}\mathbf{H}_{\nu'}\mathbf{x}_{\nu'}
  + \mathbf{H}_\nu^{\mathrm T}\mathbf{g}_z.
  \label{eq:qct_proj_general}
\end{equation}
By construction of the QCT matrices (shown in the next subsubsection), the inter-stream coupling terms vanish,
\begin{equation}
  \mathbf{H}_\nu^{\mathrm T}\mathbf{G}\mathbf{H}_{\nu'}=\mathbf{0},
  \qquad \nu'\neq \nu,
  \label{eq:qct_cross_zero_app}
\end{equation}
and the effective per-stream channel matrices are diagonal,
\begin{equation}
  \boldsymbol{\Lambda}_\nu
  \triangleq
  \mathbf{H}_\nu^{\mathrm T}\mathbf{G}\mathbf{H}_{\nu}
  =
  \operatorname{diag}(\Lambda_{\nu,0},\Lambda_{\nu,1},\ldots,\Lambda_{\nu,N_0-1}).
  \label{eq:qct_diag_app}
\end{equation}
Since $\mathbf{G}=\mathbf{C}^{\mathrm T}\mathbf{C}$, \eqref{eq:qct_cross_zero_app}-\eqref{eq:qct_diag_app} coincide with
\eqref{eq:Lambdaqctzero}-\eqref{eq:Lambdaqct} in the system model. Hence \eqref{eq:qct_proj_general} reduces to the decoupled stream
model
\begin{equation}
  \widetilde{\mathbf{x}}_{\nu}
  =
  \boldsymbol{\Lambda}_\nu \mathbf{x}_{\nu}
  + \boldsymbol{\eta}_\nu,
  \qquad
  \boldsymbol{\eta}_\nu =\mathbf{H}_\nu^{\mathrm T}\mathbf{g}_z.
  \label{eq:qct_stream_model}
\end{equation}
Moreover,
$\mathbb{E}[\boldsymbol{\eta}_\nu\boldsymbol{\eta}_\nu^{\mathrm T}]
=\sigma_g^2\mathbf{H}_\nu^{\mathrm T}\mathbf{G}\mathbf{H}_\nu=\sigma_g^2\boldsymbol{\Lambda}_\nu$ is diagonal; therefore the entries of
$\widetilde{\mathbf{x}}_{\nu}$ decouple element-wise:
\begin{equation}
  \widetilde{x}_{\nu,r} = \Lambda_{\nu,r} x_{\nu,r} + \eta_{\nu,r},
  \qquad r=0,1,\ldots,N_0-1.
  \label{eq:qct_per_entry}
\end{equation}
Consequently, for $\Lambda_{\nu,r}\neq 0$, single-tap (ZF) equalization is given by
\begin{equation}
  \widehat{x}_{\nu,r}=\frac{\widetilde{x}_{\nu,r}}{\Lambda_{\nu,r}},
  \qquad \nu\in\{1,2,3,4\},\ r=0,1,\ldots,N_0-1.
  \label{eq:qct_zf}
\end{equation}

\subsubsection{Symmetry-driven construction of QCT matrices from the $(\mathbf{J},\mathbf{S})$ invariances of $\mathbf{G}$}
\label{app:qct_design_from_symmetry}

Let $\mathbf{G}= \mathbf{C}^{\mathrm T}\mathbf{C}\in\mathbb{R}^{N\times N}$ denote the matched-filtered channel operator.
The QCT matrices are constructed by (i) exploiting two commuting permutation symmetries preserved by $\mathbf{G}$ to obtain four pairwise
orthogonal $\mathbf{G}$-invariant subspaces (which eliminates inter-stream coupling), and (ii) diagonalizing the restriction of
$\mathbf{G}$ to each subspace to enable single-tap equalization.

\paragraph{Symmetry operators.}
Assume $N$ is even and index vectors $\mathbf{u}\in\mathbb{R}^N$ by $n=0,\ldots,N-1$. Define the $N\times N$ permutation matrices
$\mathbf{J}$ (\emph{time reversal}) by
\begin{equation}
  (\mathbf{J}\mathbf{u})[n]=\mathbf{u}[N-1-n]
  \label{eq:J_defs_app_rewrite}
\end{equation}
and $\mathbf{S}$ (\emph{half-period circular shift}) by
\begin{equation}
  (\mathbf{S}\mathbf{u})[n]=\mathbf{u}\!\bigl[(n+N/2)\bmod N\bigr].
  \label{eq:S_defs_app_rewrite}
\end{equation}
Both are symmetric involutions, i.e., $\mathbf{J}^{\mathrm T}=\mathbf{J}$, $\mathbf{S}^{\mathrm T}=\mathbf{S}$,
$\mathbf{J}^2=\mathbf{S}^2=\mathbf{I}_N$, and for even $N$ they commute: $\mathbf{J}\mathbf{S}=\mathbf{S}\mathbf{J}$.

\begin{lemma}[Commutation of $\mathbf{G}$ with $(\mathbf{J},\mathbf{S})$]
\label{lem:G_commutes_JS_app}
For $\mathbf{G}=\mathbf{C}^{\mathrm T}\mathbf{C}$ with circulant $\mathbf{C}$, one has
\begin{equation}
  \mathbf{G}\mathbf{S}=\mathbf{S}\mathbf{G},
  \qquad
  \mathbf{G}\mathbf{J}=\mathbf{J}\mathbf{G}.
  \label{eq:G_commutes_app_rewrite}
\end{equation}
\end{lemma}
\begin{proof}
Since $\mathbf{G}$ is circulant, it commutes with any circular shift matrix; in particular, $\mathbf{G}\mathbf{S}=\mathbf{S}\mathbf{G}$.
Moreover, for any circulant $\mathbf{A}$ it holds that $\mathbf{J}\mathbf{A}\mathbf{J}=\mathbf{A}^{\mathrm T}$, and hence for
symmetric $\mathbf{G}$ one has $\mathbf{J}\mathbf{G}\mathbf{J}=\mathbf{G}$, which is equivalent to $\mathbf{G}\mathbf{J}=\mathbf{J}\mathbf{G}$.
\end{proof}

\paragraph{Joint eigenspace decomposition and invariance.}
Because $\mathbf{J}$ and $\mathbf{S}$ are commuting symmetric involutions, $\mathbb{R}^N$ decomposes into the orthogonal direct sum
of their four joint eigenspaces:
\begin{equation}
  \mathcal{U}_{\epsilon,\sigma}
  \triangleq
  \left\{\mathbf{u}\in\mathbb{R}^N:\ \mathbf{J}\mathbf{u}=\epsilon\,\mathbf{u},\ \mathbf{S}\mathbf{u}=\sigma\,\mathbf{u}\right\}
  \label{eq:joint_spaces_app_rewrite}
\end{equation}
where $(\epsilon,\sigma)\in\{+1,-1\}^2$. Equivalently, the orthogonal projectors onto these subspaces are
\begin{equation}
  \mathbf{P}_{\epsilon,\sigma}\triangleq \frac{1}{4}\bigl(\mathbf{I}_N+\epsilon\mathbf{J}\bigr)\bigl(\mathbf{I}_N+\sigma\mathbf{S}\bigr),
  \label{eq:projectors_app_rewrite}
\end{equation}
which satisfy $\sum_{\epsilon,\sigma}\mathbf{P}_{\epsilon,\sigma}=\mathbf{I}_N$ and
$\mathbf{P}_{\epsilon,\sigma}\mathbf{P}_{\epsilon',\sigma'}=\mathbf{0}$ whenever $(\epsilon,\sigma)\neq(\epsilon',\sigma')$.
By Lemma~\ref{lem:G_commutes_JS_app}, $\mathbf{G}$ commutes with each $\mathbf{P}_{\epsilon,\sigma}$ and therefore leaves every
$\mathcal{U}_{\epsilon,\sigma}$ invariant:
\begin{equation}
  \mathbf{G}\,\mathcal{U}_{\epsilon,\sigma}\subseteq \mathcal{U}_{\epsilon,\sigma}.
\end{equation}
This invariance is the design principle: assigning distinct data streams to distinct $\mathcal{U}_{\epsilon,\sigma}$ precludes
inter-stream coupling in the matched-filtered model \eqref{eq:qct_mf_model2_app}.

\paragraph{Constructive subspace bases and channel-independent synthesis matrices.}
Assume $N=4N_0$ and partition any $\mathbf{u}\in\mathbb{R}^N$ as
$\mathbf{u}=[\mathbf{u}_0^{\mathrm T},\mathbf{u}_1^{\mathrm T},\mathbf{u}_2^{\mathrm T},\mathbf{u}_3^{\mathrm T}]^{\mathrm T}$
with $\mathbf{u}_i\in\mathbb{R}^{N_0}$. Let $\mathbf{K}\in\mathbb{R}^{N_0\times N_0}$ denote the reversal matrix
$(\mathbf{K}\mathbf{v})[n]=\mathbf{v}[N_0-1-n]$. Then the actions of $\mathbf{S}$ and $\mathbf{J}$ admit the block forms
\begin{equation}
  \mathbf{S}\mathbf{u}=\begin{bmatrix}\mathbf{u}_2\\ \mathbf{u}_3\\ \mathbf{u}_0\\ \mathbf{u}_1\end{bmatrix},
  \qquad
  \mathbf{J}\mathbf{u}=\begin{bmatrix}\mathbf{K}\mathbf{u}_3\\ \mathbf{K}\mathbf{u}_2\\ \mathbf{K}\mathbf{u}_1\\ \mathbf{K}\mathbf{u}_0\end{bmatrix}.
  \label{eq:JS_block_actions_app_rewrite}
\end{equation}
Imposing $\mathbf{S}\mathbf{u}=\sigma\mathbf{u}$ and $\mathbf{J}\mathbf{u}=\epsilon\mathbf{u}$ yields the block relations
\begin{equation}
  \mathbf{u}_2=\sigma\mathbf{u}_0,\qquad
  \mathbf{u}_1=\epsilon\sigma\,\mathbf{K}\mathbf{u}_0,\qquad
  \mathbf{u}_3=\epsilon\,\mathbf{K}\mathbf{u}_0,
  \label{eq:block_relations_app_rewrite}
\end{equation}
and hence the parametrization
\begin{equation}
  \mathbf{u}
  =\frac{1}{2}\begin{bmatrix}
    \mathbf{v}\\ \epsilon\sigma\,\mathbf{K}\mathbf{v}\\ \sigma\,\mathbf{v}\\ \epsilon\,\mathbf{K}\mathbf{v}
  \end{bmatrix},
  \qquad \mathbf{v}\in\mathbb{R}^{N_0}.
  \label{eq:U_parametrization_app_rewrite}
\end{equation}
Consequently, for any orthonormal $\mathbf{B}\in\mathbb{R}^{N_0\times N_0}$, the matrix
\begin{equation}
  \boldsymbol{\Gamma}_{\epsilon,\sigma}(\mathbf{B})
  \triangleq
  \frac{1}{2}\begin{bmatrix}
    \mathbf{B}\\ \epsilon\sigma\,\mathbf{K}\mathbf{B}\\ \sigma\,\mathbf{B}\\ \epsilon\,\mathbf{K}\mathbf{B}
  \end{bmatrix}
  \in\mathbb{R}^{N\times N_0}
  \label{eq:Gamma_general_app_rewrite}
\end{equation}
has orthonormal columns spanning $\mathcal{U}_{\epsilon,\sigma}$. 
In particular, using an orthonormal DCT-II matrix $\boldsymbol{\Theta}=(\theta_{p,q})_{0\le p,q\le N_0-1}$ and an orthonormal DST-II matrix
$\boldsymbol{\Psi}=(\psi_{p,q})_{0\le p,q\le N_0-1}$, the channel independent synthesis matrices are
\begin{subequations}\label{eq:Gamma_defs_app_rewrite}
\begin{align}
  \boldsymbol{\Gamma}_1 &= \boldsymbol{\Gamma}_{+,+}(\boldsymbol{\Theta})
  = \frac{1}{2}\!\begin{bmatrix}\boldsymbol{\Theta}\\ \mathbf{K}\boldsymbol{\Theta}\\ \boldsymbol{\Theta}\\ \mathbf{K}\boldsymbol{\Theta}\end{bmatrix},\\
  \boldsymbol{\Gamma}_2 &= \boldsymbol{\Gamma}_{-,+}(\boldsymbol{\Psi})
  = \frac{1}{2}\!\begin{bmatrix}\boldsymbol{\Psi}\\ -\mathbf{K}\boldsymbol{\Psi}\\ \boldsymbol{\Psi}\\ -\mathbf{K}\boldsymbol{\Psi}\end{bmatrix},\\
  \boldsymbol{\Gamma}_3 &= \boldsymbol{\Gamma}_{-,-}(\boldsymbol{\Theta})
  = \frac{1}{2}\!\begin{bmatrix}\boldsymbol{\Theta}\\ \mathbf{K}\boldsymbol{\Theta}\\ -\boldsymbol{\Theta}\\ -\mathbf{K}\boldsymbol{\Theta}\end{bmatrix},\\
  \boldsymbol{\Gamma}_4 &= \boldsymbol{\Gamma}_{+,-}(\boldsymbol{\Psi})
  = \frac{1}{2}\!\begin{bmatrix}\boldsymbol{\Psi}\\ -\mathbf{K}\boldsymbol{\Psi}\\ -\boldsymbol{\Psi}\\ \mathbf{K}\boldsymbol{\Psi}\end{bmatrix}.
\end{align}
\end{subequations}
The DCT-II and DST-II matrices used in the implementation are orthonormal and indexed by
$p,q\in\{0,1,\ldots,N_0-1\}$. Their entries are, respectively,
\begin{equation}
\theta_{p,q}=
\begin{cases}
\sqrt{\frac{1}{N_0}}, & p=0,\\[0.5ex]
\sqrt{\frac{2}{N_0}}\cos\!\left(\frac{\pi(2q+1)p}{2N_0}\right), & p=1,2,\ldots,N_0-1,
\end{cases}
\label{eq:DCT_entries}
\end{equation}
and
\begin{equation}
\psi_{p,q}=\alpha_p\,
\sin\!\left(\frac{\pi(2q+1)(p+1)}{2N_0}\right),
\label{eq:DST_entries}
\end{equation}
where
\begin{equation}
\alpha_p=
\begin{cases}
\sqrt{\frac{2}{N_0}}, & p=0,1,\ldots,N_0-2,\\[0.5ex]
\sqrt{\frac{1}{N_0}}, & p=N_0-1.
\end{cases}
\label{eq:DST_norm}
\end{equation}

\begin{lemma}[Inter-stream decoupling under $\mathbf{G}$]
\label{lem:Gamma_blockdiag_app}
For $\nu\neq\nu'$, the cross terms vanish:
\begin{equation}
  \boldsymbol{\Gamma}_\nu^{\mathrm T}\mathbf{G}\boldsymbol{\Gamma}_{\nu'}=\mathbf{0}.
  \label{eq:Gamma_cross_zero_app_rewrite}
\end{equation}
\end{lemma}
\begin{proof}
The ranges of $\boldsymbol{\Gamma}_\nu$ and $\boldsymbol{\Gamma}_{\nu'}$ are contained in distinct joint eigenspaces
$\mathcal{U}_{\epsilon,\sigma}$ and $\mathcal{U}_{\epsilon',\sigma'}$ with $(\epsilon,\sigma)\neq(\epsilon',\sigma')$.
These subspaces are orthogonal and $\mathbf{G}$ invariant; hence $\mathbf{G}\boldsymbol{\Gamma}_{\nu'}$ lies in
$\mathcal{U}_{\epsilon',\sigma'}$ and is orthogonal to $\mathcal{U}_{\epsilon,\sigma}$, implying
$\boldsymbol{\Gamma}_\nu^{\mathrm T}\mathbf{G}\boldsymbol{\Gamma}_{\nu'}=\mathbf{0}$.
\end{proof}

\paragraph{Single-tap diagonalization within each subspace.}
While $\boldsymbol{\Gamma}_\nu$ guarantees inter-stream decoupling, $\mathbf{G}$ may act non-diagonally within each
$\mathcal{U}_{\epsilon,\sigma}$. Define the reduced symmetric blocks
\begin{equation}
  \mathbf{\Omega}_\nu \triangleq \boldsymbol{\Gamma}_\nu^{\mathrm T}\mathbf{G}\boldsymbol{\Gamma}_\nu
  \in\mathbb{R}^{N_0\times N_0},\qquad \nu\in\{1,2,3,4\}.
  \label{eq:Omega_def_app_rewrite}
\end{equation}
Let $\mathbf{\Omega}_\nu=\mathbf{D}_\nu\boldsymbol{\Lambda}_\nu\mathbf{D}_\nu^{\mathrm T}$ denote an eigen-decomposition with
orthonormal $\mathbf{D}_\nu$ and diagonal $\boldsymbol{\Lambda}_\nu\succeq \mathbf{0}$, and define the final QCT matrices as
\begin{equation}
  \mathbf{H}_\nu \triangleq \boldsymbol{\Gamma}_\nu\mathbf{D}_\nu \in \mathbb{R}^{N\times N_0}.
  \label{eq:H_def_app_rewrite}
\end{equation}

Then Lemma~\ref{lem:Gamma_blockdiag_app} and \eqref{eq:Omega_def_app_rewrite} imply that, for $\nu\neq \nu'$,
the cross terms vanish and the within stream blocks diagonalize under $\mathbf{G}$. Equivalently, one obtains the inter‑stream
orthogonality and per‑stream diagonalization stated in \eqref{eq:Lambdaqctzero} and \eqref{eq:Lambdaqct}, respectively. Accordingly,
after matched filtering and projection by $\mathbf{H}_\nu^{\mathrm T}$, each stream admits scalar (single‑tap) equalization by
element‑wise division with $\operatorname{diag}(\boldsymbol{\Lambda}_\nu)$.

\section{Optical, Electrical, and Clipping Power}
\label{app:gaussian_clipping_moments}
\subsection{Moments of a DC-Biased Half-Wave Clipped Gaussian Sample}
\label{app:clipping_moments}

\paragraph{Reminder}
Let $X\sim\mathcal{N}(0,\sigma^2)$ with $\sigma^2>0$. For a electrical bias $B_{\mathrm{DC}}\ge 0$, define the
biased and half-wave clipped sample

\begin{equation}
  Y \triangleq \max\{X+B_{\mathrm{DC}},\,0\}=(X+B_{\mathrm{DC}})\,\mathbbm{1}_{\{X\ge -B_{\mathrm{DC}}\}}.
  \label{eq:Y_def_app}
\end{equation}
The associated clipping term is
\begin{equation}
  c \triangleq Y-(X+B_{\mathrm{DC}})=-(X+B_{\mathrm{DC}})\,\mathbbm{1}_{\{X<-B_{\mathrm{DC}}\}}\ \ 
  \label{eq:c_def_app}
\end{equation}
where $  B_{\mathrm{DC}} \triangleq \mu\,\sigma_{x}$.

Let $Z\sim\mathcal{N}(0,1)$ with pdf $f_Z$ and cdf $F_Z$:

\begin{subequations}\label{eq:tail_identities_app}
\begin{align}
  f_Z(z)\triangleq \frac{1}{\sqrt{2\pi}}e^{-z^2/2}, \label{eq:f_z}\\
  F_Z(z)\triangleq \int_{-\infty}^{z} f_Z(t)\,\mathrm{d}t,\label{eq:F_z}\\
  \bar F_Z(z)\triangleq 1-F_Z(z)=\int_{z}^{\infty} f_Z(t)\,\mathrm{d}t. \label{eq:Fbar_z}
\end{align}
\end{subequations}
Note that $f_Z(-z)=f_Z(z)$ and $F_Z(-z)=\bar F_Z(z)$. For any $a\in\mathbb{R}$,
\begin{subequations}\label{eq:tail_identities_app2}
\begin{align}
  \int_{a}^{\infty} f_Z(z)\,\mathrm{d}z &= \bar F_Z(a), \label{eq:tail0_app}\\
  \int_{a}^{\infty} z\, f_Z(z)\,\mathrm{d}z &= f_Z(a), \label{eq:tail1_app}\\
  \int_{a}^{\infty} z^2\, f_Z(z)\,\mathrm{d}z &= \bar F_Z(a)+a f_Z(a). \label{eq:tail2_app}
\end{align}
\end{subequations}
The identities \eqref{eq:tail1_app}-\eqref{eq:tail2_app} follow from $f_Z'(z)=-z f_Z(z)$ and integration by parts.

\paragraph{Mean optical (average) power: $\mathbb{E}[Y]$.}
Using \eqref{eq:Y_def_app} and the change of variables $x=\sigma z$ (so that $z=-\mu$ corresponds to $x=-B_{\mathrm{DC}}$),
\begin{align}
  \mathbb{E}[Y]
  &= \int_{-B_{\mathrm{DC}}}^{\infty} (x+B_{\mathrm{DC}})\, f_X(x)\,\mathrm{d}x \nonumber\\
   &= \int_{-\mu}^{\infty} (\sigma z+B_{\mathrm{DC}})\, f_Z(z)\,\mathrm{d}z \\
  &= \sigma \int_{-\mu}^{\infty} z f_Z(z)\,\mathrm{d}z + B_{\mathrm{DC}}\int_{-\mu}^{\infty} f_Z(z)\,\mathrm{d}z.\nonumber
\end{align}
By symmetry and \eqref{eq:tail1_app}-\eqref{eq:tail0_app},
\[
\int_{-\mu}^{\infty} z f_Z(z)\,\mathrm{d}z=f_Z(\mu),
\qquad
\int_{-\mu}^{\infty} f_Z(z)\,\mathrm{d}z=F_Z(\mu),
\]
hence
\begin{equation}
   \mathbb{E}[Y]=\sigma\,f_Z(\mu)+B_{\mathrm{DC}}\,F_Z(\mu).\ 
  \label{eq:EY_app}
\end{equation}

\paragraph{Electrical power: $\mathbb{E}[Y^2]$.}
Similarly, since $Y^2=(X+B_{\mathrm{DC}})^2\,\mathbbm{1}_{\{X\ge -B_{\mathrm{DC}}\}}$,
\begin{align}
  \mathbb{E}[Y^2]
  &=\int_{-\mu}^{\infty}(\sigma z+B_{\mathrm{DC}})^2 f_Z(z)\,\mathrm{d}z \nonumber\\
  &=\sigma^2\!\int_{-\mu}^{\infty}\! z^2 f_Z(z)\,\mathrm{d}z
   +2\sigma B_{\mathrm{DC}}\!\int_{-\mu}^{\infty}\! z f_Z(z)\,\mathrm{d}z \\
   &+B_{\mathrm{DC}}^2\!\int_{-\mu}^{\infty}\! f_Z(z)\,\mathrm{d}z\nonumber.
\end{align}
Using $\int_{-\mu}^{\infty} z^2 f_Z(z)\,\mathrm{d}z = F_Z(\mu)-\mu f_Z(\mu)$ along with the previous two integrals gives
\begin{equation}
  \ \mathbb{E}[Y^2]=(\sigma^2+B_{\mathrm{DC}}^2)\,F_Z(\mu)+\sigma B_{\mathrm{DC}}\,f_Z(\mu).\ 
  \label{eq:EY2_app}
\end{equation}

\paragraph{Clipping power: $\mathbb{E}[c^2]$.}
From \eqref{eq:c_def_app}, we have $c^2=(X+B_{\mathrm{DC}})^2\,\mathbbm{1}_{\{X<-B_{\mathrm{DC}}\}}$ and thus
\begin{align}
  \mathbb{E}[c^2]
  &=\int_{-\infty}^{-\mu}(\sigma z+B_{\mathrm{DC}})^2 f_Z(z)\,\mathrm{d}z \nonumber\\
  &= (\sigma^2+B_{\mathrm{DC}}^2)\,\bar F_Z(\mu) - \sigma B_{\mathrm{DC}}\, f_Z(\mu) \nonumber\\
  &= \sigma^2\Big((1+\mu^2)\,\bar F_Z(\mu)-\mu f_Z(\mu)\Big).
\end{align}
Therefore,

\begin{align}
   \mathbb{E}[c^2]&=(\sigma^2+B_{\mathrm{DC}}^2)\,\bar F_Z(\mu)-\sigma B_{\mathrm{DC}}\,f_Z(\mu) \nonumber\\
  &= \sigma^2\!\left((1+\mu^2)\bar F_Z(\mu)-\mu f_Z(\mu)\right).\ 
  \label{eq:Ec2_app}
\end{align}

Since $(X+B_{\mathrm{DC}})=Y-c$ with $Yc=0$ almost surely, one obtains
\begin{equation}
  \mathbb{E}[Y^2]+\mathbb{E}[c^2]=\mathbb{E}[(X+B_{\mathrm{DC}})^2]=\sigma^2+B_{\mathrm{DC}}^2,
  \label{eq:power_identity_app}
\end{equation}
which provides a useful numerical check for Monte-Carlo simulations.

\paragraph{Identity, limits, and numerical sanity checks.}
The identity in \eqref{eq:power_identity_app} follows from the orthogonal decomposition
$(X+B_{\mathrm{DC}})=Y-c$ with $Yc=0$ almost surely (only one of $Y$ and $c$ is nonzero for any realization). Equivalently, in terms of the
normalized bias $\mu\triangleq B_{\mathrm{DC}}/\sigma$, one may write
\begin{equation}
  \mathbb{E}[Y^{2}] + \mathbb{E}[c^{2}]
  = \sigma^{2} + B_{\mathrm{DC}}^{2}
  = \sigma^{2}\bigl(1+\mu^{2}\bigr),
  \label{eq:power_identity_mu_form_app}
\end{equation}
which is independent of the clipping probability. The latter is
\begin{equation}
  p_{\mathrm{clip}}
  \triangleq \Pr\{X<-B_{\mathrm{DC}}\}
  = \bar{F}_{Z}(\mu).
  \label{eq:pclip_app}
\end{equation}
Two useful limiting cases are:
\begin{itemize}\setlength{\itemsep}{0.15em}
  \item \emph{No bias} ($\mu=0$, i.e., $B_{\mathrm{DC}}=0$): $Y=\max\{X,0\}$ (half-wave rectification) and
  $\mathbb{E}[Y]=\sigma/\sqrt{2\pi}$, $\mathbb{E}[Y^{2}]=\sigma^{2}/2$, $\mathbb{E}[c^{2}]=\sigma^{2}/2$.
  \item \emph{Large bias} ($\mu\to\infty$): $p_{\mathrm{clip}}\to 0$ and $Y\to X+B_{\mathrm{DC}}$, hence
  $\mathbb{E}[Y]\to B_{\mathrm{DC}}$ (equivalently, $\mathbb{E}[Y]/\sigma\to \mu$),
  $\mathbb{E}[Y^{2}]\to \sigma^{2}+B_{\mathrm{DC}}^{2}$, and $\mathbb{E}[c^{2}]\to 0$.
\end{itemize}
In simulations, we monitor the relative mismatch of \eqref{eq:power_identity_app} as
\begin{equation}
  \epsilon_{\mathrm{rel}}
  \triangleq
  \frac{\bigl|\bigl(\mathbb{E}[Y^{2}]+\mathbb{E}[c^{2}]\bigr)-\bigl(\sigma^{2}+B_{\mathrm{DC}}^{2}\bigr)\bigr|}
       {\sigma^{2}+B_{\mathrm{DC}}^{2}},
  \label{eq:relerr_identity_app}
\end{equation}
which should be close to numerical precision when expectations are estimated accurately.

\paragraph{Mapping to DCO-OFDM and QCT-OFDM.}
For DCO-OFDM, interpret $X$ as the unbiased bipolar OFDM time sample (per equivalent color branch) with standard deviation $\sigma_x$ and bias
$B_{\mathrm{DC}}=\mu\sigma_x$ (cf.\ \eqref{eq:dc_bias_mu}); then $P_{\mathrm{opt}}=\mathbb{E}[Y]$, $P_{\mathrm{elec}}=\mathbb{E}[Y^{2}]$, and
$P_{\mathrm{clip}}=\mathbb{E}[c^{2}]$ follow directly from \eqref{eq:EY_app}--\eqref{eq:Ec2_app}.
For QCT-OFDM, the same formulas apply per branch $\nu$ after the substitutions
$(\sigma_x,B_{\mathrm{DC}})\mapsto (\sigma_{x,\nu},B_{\mathrm{DC},\nu})$.
If multiple co-located branches/LEDs are summed optically, the corresponding mean powers add linearly across branches.

\paragraph{EVM computation.}
For each DCO-OFDM block, the frequency-domain OFDM vector $\mathbf{x}\in\mathbb{C}^{N}$ is formed with Hermitian symmetry as in
\eqref{eq:HermitianX}, and the real bipolar time-domain block is obtained by the unitary IDFT
$\mathbf{x}_{f}=\mathbf{F}^{\mathrm H}\mathbf{x}$ in \eqref{eq:ifft_output}. After DC biasing and half-wave clipping
(\eqref{eq:dc_bias_add}--\eqref{eq:dc_bias_clip}), CP insertion/removal, and propagation through the circulant channel
$\mathbf{C}$, the receiver obtains $\mathbf{y}\in\mathbb{R}^{N}$.
Prior to FFT demodulation, the deterministic bias contribution is removed using the known bias term, yielding
\begin{equation}
\mathbf{y}_{0}\triangleq \mathbf{y}-\mathbf{C}\bigl(B_{\mathrm{DC}}\mathbf{1}_{N}\bigr),
\label{eq:rx_bias_remove_evm}
\end{equation}
so that the subsequent FFT primarily reflects the unbiased OFDM component plus clipping distortion and noise.
Applying the FFT gives $\tilde{\mathbf{x}}=\mathbf{F}\mathbf{y}_{0}$, and single-tap ZF equalization is then performed on each
data subcarrier $k=p+1\in\mathcal{K}_{\mathrm{data}}$ as in \eqref{eq:s_hat}, i.e.,
$\tilde{s}_{p}=\tilde{x}_{p+1}/\Lambda_{p+1}$ for $p=0,1,\ldots,\frac{N}{2}-2$, where $\Lambda_{k}$ denotes the $k$th
frequency-domain channel gain (Appendix~\ref{app:ofdm_single_tap}). Let
$\mathbf{s}=[s_{0},\ldots,s_{N/2-2}]^{\mathsf T}$ and
$\tilde{\mathbf{s}}=[\tilde{s}_{0},\ldots,\tilde{s}_{N/2-2}]^{\mathsf T}$ denote the transmitted and equalized data-symbol vectors,
respectively. The (per-block) normalized mean-square EVM is computed over the data subcarriers as
\begin{equation}
\mathrm{EVM}^{2}
\triangleq
\frac{\left\lVert \tilde{\mathbf{s}}-\mathbf{s}\right\rVert_{2}^{2}}
{\left\lVert \mathbf{s}\right\rVert_{2}^{2}},
\label{eq:evm_def_block}
\end{equation}
and the reported RMS EVM at a given SNR point is obtained by averaging \eqref{eq:evm_def_block} over $N_{\mathrm{blk}}$ Monte-Carlo blocks and
taking the square root,
\begin{equation}
\mathrm{EVM}_{\mathrm{tot}}[\%]
=
100\sqrt{\frac{1}{N_{\mathrm{blk}}}\sum_{b=1}^{N_{\mathrm{blk}}}
\frac{\left\lVert \tilde{\mathbf{s}}^{(b)}-\mathbf{s}^{(b)}\right\rVert_{2}^{2}}
{\left\lVert \mathbf{s}^{(b)}\right\rVert_{2}^{2}} }.
\label{eq:evm_tot_mc}
\end{equation}
To isolate nonlinear distortion due to transmitter clipping, the clipping-only EVM, $\mathrm{EVM}_{\mathrm{clip}}$, is computed by repeating
the same receiver processing (bias removal, FFT, and ZF equalization) with the AWGN disabled (i.e., $\mathbf{g}=\mathbf{0}$ in the received-block
model), and then evaluating \eqref{eq:evm_def_block}--\eqref{eq:evm_tot_mc}. In this way, $\mathrm{EVM}_{\mathrm{clip}}$ captures the distortion floor
induced solely by the half-wave rectifier, whereas $\mathrm{EVM}_{\mathrm{tot}}$ reflects the combined impact of clipping distortion, channel effects,
and AWGN under the SNR definition in Sec.~\ref{sec:fairness}.

\section*{Acknowledgment}
This work was supported by the Scientific and Technological Research Council of Türkiye (T\"UB\.ITAK) under Grant Project~125E354.

\bibliographystyle{IEEEtran}  
\bibliography{ref}       

\end{document}